\documentclass[usegraphicx,useAMS]{mn2e}
\usepackage{multicol,lscape}
\input{psfig.sty}

\newcommand{\AIPS}{{$\cal AIPS\/$}}

\newcommand{\Smed}{$S_{\rm 1,400}^{\rm med}$}
\newcommand{\Smean}{$\bar{S}_{\rm 1,400}$}
\newcommand{\Sgmrt}{$\bar{S}_{\rm 610}$}
\newcommand{\Smedgmrt}{$S_{\rm 610}^{\rm med}$}
\newcommand{\Sint}{$S_{\rm 1,400}^{\rm int}$}
\newcommand{\Srad}{$S_{\rm 1,400}$}
\newcommand{\Lrad}{$L_{\rm 1,400}$}

\def\gs{\mathrel{\raise0.35ex\hbox{$\scriptstyle >$}\kern-0.6em
\lower0.40ex\hbox{{$\scriptstyle \sim$}}}}
\def\ls{\mathrel{\raise0.35ex\hbox{$\scriptstyle <$}\kern-0.6em
\lower0.40ex\hbox{{$\scriptstyle \sim$}}}}

\title[The star-formation history of $K$-selected galaxies]
      {The star-formation history of $K$-selected galaxies}
 
\author[L.\ Dunne et al.]
{L.\ Dunne,$^{\! 1}$
R.\,J.\ Ivison,$^{\! 2,3}$
S.\ Maddox,$^{\! 1}$
M.\ Cirasuolo,$^{\! 3}$
A.\,M.\ Mortier,$^{\! 3}$
S.\ Foucaud,$^{\! 1}$
\and
E.\ Ibar,$^{\! 2}$
O.\ Almaini,$^{\! 1}$
C.\ Simpson$^{4}$ and
R.\ McLure$^{3}$
\vspace*{1mm}\\
$^{1}$ School of Physics and Astronomy, University of Nottingham,
       University Park, Nottingham NG7 2RD\\
$^{2}$ UK Astronomy Technology Centre, Royal Observatory, Blackford
       Hill, Edinburgh EH9 3HJ\\
$^{3}$ Institute for Astronomy, University of Edinburgh, Royal Observatory, Blackford
       Hill, Edinburgh EH9 3HJ\\
$^{4}$ Astrophysics Research Institute, Liverpool John Moores
       University, Twelve Quays House, Egerton Wharf, Birkenhead CH41 1LD}
\date{Accepted 2008 August 22. Received 2008 August 21; in original form 2008 May 25}

\volume{390}
\pubyear{2009}
\pagerange{3--20}

\begin{document}
 
\maketitle

\begin{abstract}
We have studied the $\mu$Jy radio properties of $K$-selected galaxy
populations detected in the Ultra-Deep Survey (UDS) portion of the
United Kingdom Infrared Telescope (UKIRT) Deep Sky Survey (UKIDSS)
using 610- and 1,400-MHz images from the Very Large Array (VLA) and
the Giant Metre-wave Telescope (GMRT). These deep radio mosaics,
combined with the largest and deepest $K$-band image currently
available, allow high-signal-to-noise (S/N) detections of many
$K$-selected sub-populations, including sBzK and pBzK star-forming and
passive galaxies. We find a strong correlation between the radio and
$K$-band fluxes and a linear relationship between star-formation rate
(SFR) and $K$-band luminosity. We find no evidence, from either radio
spectral indices or a comparison with submm-derived SFRs, that the
full sample is strongly contaminated by active galactic nuclei (AGN)
at these low flux densities, though this is very difficult to
determine from this dataset. The photometric redshift distributions
for the BzK galaxies place 37 (29) per cent of pBzK (sBzK) galaxies at
$z<1.4$, implying that location on the BzK diagram alone is not
sufficient to select samples at $1.4<z<2.5$. The sBzK and pBzK
galaxies have similar levels of radio flux density, SFR and specific
SFR (SSFR -- SFR per unit stellar mass) at $z<1.4$, suggesting there
is strong contamination of the pBzK sample by star-forming
galaxies. At $z>1.4$, the pBzK galaxies become difficult to detect in
the radio stack, though the implied SFRs are still much higher than
expected for passively evolving galaxies. It may be that their radio
emission comes from low-luminosity AGN. Extremely red objects (EROs)
straddle the passive and star-forming regions of the BzK diagram and
also straddle the two groups in terms of their radio properties. We
find that $K$-bright ERO samples are dominated by passive galaxies and
faint ERO samples contain more star-forming galaxies.  The
star-formation history (SFH) from stacking all $K$-band sources in the
UDS agrees well with that derived for other wavebands and other radio
surveys, at least out to $z\sim 2$. The radio-derived SFH then appears
to fall more steeply than that measured at other wavelengths. The SSFR
for $K$-selected sources rises strongly with redshift at all stellar
masses, and shows a weak dependence on stellar mass. High- and
low-mass galaxies show a similar decline in SSFR since $z\sim 2$.
\end{abstract}

\begin{keywords}
galaxies: active -- galaxies: evolution -- galaxies: starburst
\end{keywords}

\section{Introduction}

There has recently been a large advance in our ability to undertake
deep, large-area surveys in the near-infrared (IR) due to the
introduction of very large IR-sensitive arrays. Such surveys allow us
to collect statistically powerful samples of galaxies selected in the
$K$ filter at 2.2\,$\mu$m, providing access to the part of the
spectral energy distribution (SED) dominated by rest-frame optical
light for galaxies at high redshift and by low-mass stars for galaxies
at low redshift. Since it is not as strongly affected by dust or
biased towards young stars as optical emission, it provides a sample
selected approximately on stellar mass ($M_{\rm stellar}$).

There are several classifications for $K$-selected galaxies --
extremely red objects (EROs -- Elston, Rieke \& Rieke 1989), distant
red galaxies (DRGs -- van Dokkum et al.\ 2004) and, more recently, BzK
galaxies (Daddi et al.\ 2004). All these selections are colour based
and aim to select -- via SED shape -- particular kinds of galaxies at
particular redshifts. The most successful selection appears to be via
the BzK technique, where both passive (pBzK) and star-forming (sBzK)
galaxies can be selected over the interval $1.4<z<2.5$. In order to
investigate the star-formation properties of these galaxies, we will
exploit deep radio imaging and stack the IR positions into the radio
data to derive averaged radio properties at $\mu$Jy levels for various
classes of $K$-selected sources.

The strong correlation between far-IR (FIR) and radio emission (Helou
et al.\ 1985) in local star-forming galaxies (SFGs) has allowed the
use of radio surveys as a means to quantify star-formation activity in
a manner immune to the obscuring effects of dust. The FIR emission is
believed to be produced by dust which has been heated by the
ultraviolet (UV) photons from massive stars. These stars end their
lives as supernovae and the remnants of these explosions produce the
synchrotron radio emission. Thus both the FIR and radio trace the rate
of formation of massive stars. Radio emission is unaffected by dust,
unlike optical and UV photons, and so the radio provides a powerful
tool for studying star formation at high angular resolution in a
relatively unbiased way.

The radio source counts above 1\,mJy are dominated by AGN, the
powerful Fanaroff-Riley-{\sc ii} (FR\,{\sc ii} -- Fanaroff \& Riley
1974) sources giving way to the less powerful (but still radio-loud)
FR\,{\sc i} AGN as the flux-density threshold approaches $S_{\rm
1,400}\sim 1$\,mJy (Windhorst et al.\ 1985; Willott et al.\ 2002). The
evolution of these two types of radio-emitting AGN differs, with
FR\,{\sc ii} sources undergoing strong evolution (Dunlop \& Peacock
1990; Willott et al.\ 2001) while FR\,{\sc i} sources undergo a weaker
evolution or possibly maintain a constant co-moving space density
(Clewley \& Jarvis 2004; Sadler et al.\ 2007; Rigby, Best \& Snellen
2008). The contribution of these sources to the radio source counts
should continue to be a power law to fainter flux densities (Jarvis \&
Rawlings 2004). At flux densities below 1\,mJy, the radio counts are
observed to flatten and a new population of sub-mJy radio sources,
originally thought to be SFGs, becomes apparent (Condon 1984). There
is the possibility, however, that the sub-mJy population could also
consist of `radio-quiet' AGN (Jarvis \& Rawlings 2004) -- i.e.\ AGN
with a low ratio of radio to optical/X-ray luminosity which still
contain a super-massive black hole accreting at a high rate,
contributing to the radio emission.

The exact mix of AGN versus star formation at sub-mJy levels is still
the subject of lively debate (see Ibar et al.\ 2008 and references
therein). The first deep radio surveys ascribed the bulk of the
sub-mJy activity to SFGs (Hopkins et al.\ 1998; Gruppioni et al.\
1999; Richards 2000) though further investigations with more complete
optical, X-ray and high-resolution radio imaging have shown a very
mixed picture. A deep, narrow 5-GHz survey in the Lockman Hole
(Ciliegi et al.\ 2003) found a flattening of radio spectral indices at
faint flux limits, consistent with an increasing fraction of AGN. They
also found that a substantial fraction (50 per cent) of $S_{\rm
5GHz}>50$-$\mu$Jy sources are early-type galaxies. Using radio
morphologies to distinguish radio AGN from SFGs suggests a mix of both
types at faint flux levels across a broad range of redshift and radio
luminosity (Richards et al.\ 2007), although approaching this with
sufficient resolution awaits the commissioning of e-MERLIN. Simpson et
al.\ (2006) studied the the nature of the sub-mJy population ($S_{\rm
1,400}> 100\,\mu$Jy) in the Subaru-{\em XMM-Newton} Deep Field (SXDF)
and concluded that radio-quiet AGN make a significant contribution to
the counts at $100 < S_{\rm 1,400} < 300\,\mu$Jy. This conclusion is
supported by Smol\v{c}i\'{c} et al.\ (2008) in the COSMOS field, where they
find that 50--60 per cent of $z<1.3$ galaxies with $50 <S_{\rm
1,400}<70\,\mu$Jy are AGN. In contrast, a recent study by Seymour et
al.\ (2008), using VLA and MERLIN images, has been able to separate
radio sources into AGN and SFGs on the basis of radio morphology,
spectral index and radio/near-IR and mid-IR/radio flux density
ratios. They find that SFGs dominate the counts at $\sim$50$\mu$Jy and
account for the majority of the upturn in the radio counts below
1\,mJy. An extrapolation of their figure~3 to the flux densities we
are typically probing here ($\sim 10\,\mu$Jy) would suggest that AGN
are far outnumbered by SFGs and should not be a major concern in this
stacking.


As we are probing the radio population at flux density levels an order
of magnitude fainter than existing studies, it is impossible to know
-- {\em a priori} -- the relative contributions to the radio flux of
accretion and star formation for a given population. This difficulty
in distinguishing between star-formation- and AGN-induced radio
emission will potentially present us with biases in our results. For
instance, if we assume all of the radio emission is due to star
formation, then we will over-estimate the true SFRs. If, in the hope
of excising AGN, we discount all sources with certain values of
$L_{\rm 1,400}$ or ratios of radio/optical flux, then we will bias
ourselves as extreme starbursts will also have high $L_{\rm 1,400}$
(Ivison et al.\ 2002) and high radio/optical ratios. Given the
findings of Richards et al.\ (2007), that radio-emitting AGN appear at
all radio luminosities and radio/optical ratios, we do not feel that
such cuts to the sample can be justified. We will simply include all
sources, placing an upper limit on the star-formation activity,
investigating along the way whether there are any other indications as
to the nature of the radio emission.

The layout of the paper is as follows: \S2 describes the radio and IR
observations. \S3 describes the stacking procedure used in the
radio. In \S4 we present the results of the stacks while in \S5 we
investigate the possible contamination from AGN using radio spectral
indices and submm stacks. In \S6 we present the radio-derived SFH of
$K$-selected galaxies out to $z=4$ and also investigate their specific
SFRs as a function of mass and redshift.

The cosmology used throughout is $\Omega_{\Lambda} = 0.73$,
$\Omega_m=0.27$ and $H_0 = 71$\,km\,s$^{-1}$\,Mpc$^{-1}$.

\section{Observations}
\subsection{IR data and photometric redshifts}
\label{IR-obs}

The catalogue used in this paper was selected from the ongoing UDS,
which is the deepest of the five surveys that comprise UKIDSS
(Lawrence et al.\ 2007), exploiting the IR Wide-Field Camera (WFCAM --
Casali et al.\ 2007) on UKIRT\footnote{The United Kingdom Infrared
Telescope is operated by the Joint Astronomy Centre on behalf of the
Science and Technology Facilities Council of the U.K.}. The UDS will
cover 0.8\,deg$^2$ in $J, H, K$ to eventual 5$\sigma$ limits of 26.9,
25.9, 24.9 mag (we use AB magnitudes throughout, unless stated
otherwise), built from a 4-point mosaic of WFCAM. Stacking, mosaicing
and quality control for the UDS was performed using a different recipe
to the standard UKIDSS pipeline, and is described fully in Almaini et
al.\ (in preparation). In this work we used data from the DR1 release
(Warren et al.\ 2007), which includes $J$ and $K$ images reaching
5-$\sigma$ depths of $K=23.6$ and $J=23.7$ (2-arcsec diameter
apertures). These images have median seeing of 0.85 and 0.75\,arcsec
({\sc fwhm}) in $J$ and $K$, respectively, with astrometry (calibrated
against 2MASS stars) accurate to $\sigma <0.05$\,arcsec. Photometry is
also uniformly accurate to $<$2 per cent r.m.s.\ in both bands. We
performed source extraction using the Sextractor software (Bertin \&
Arnouts 1996), using simulations to optimise source completeness,
further details of which are described in Foucaud et al.\
(2007). Objects which were saturated, in noisy regions or defined as
compact ($K<19.9$ and with a 50 per cent light radius around the point
spread function -- PSF) were not included in the catalogue. The
detection completeness at $K=23.0$ in 2-arcsec diameter apertures is
estimated to be $\sim$80 percent. At $K<22.5$ completeness is
$\sim$100 percent.

We matched the UDS $K$-selected catalogue to deep optical data in this
field from the Subaru {\em XMM-Newton\/} Deep Field survey (Furasawa
et al.\ 2008) in five broad-band filters to 3-$\sigma$ depths of
$B=28.4, V=27.8, R=27.7, i'=27.7, z'=26.6$ (again, 2-arcsec diameter
apertures).

This deep multi-wavelength coverage of the UDS allowed us to model the
galaxy SEDs in detail, as described in Cirasuolo et al.\ (2008).  The
SED fitting procedure, to derive photometric redshifts, was performed
with a code based largely on the public package {\it Hyperz}
(Bolzonella, Miralles \& Pell\'{o} 2000) and exploits reliable
photometry in 16 broad bands from far-UV to the mid-IR.  Both
empirical (Coleman, Wu \& Weeman 1980; Kinney et al.\ 1996; Mignoli et
al.\ 2005) and synthetic templates (Bruzual \& Charlot 2003) were used
to model the galaxy SEDs, including a prescription for dust
attenuation (Calzetti et al.\ 2000) and Lyman-series absorption due to
the H\,{\sc i} clouds in the intergalactic medium, according to Madau
(1995).

For the $\simeq$1,200 galaxies in the UDS with reliable spectroscopic
redshifts, the agreement of the photometric redshifts is good over the
full redshift range, $0 <z < 6$. The distribution of the $|\Delta
z|/(1+z) \equiv |(z_{\rm spect} - z_{\rm phot}) |/ (1+z_{\rm spect})$
has a mean consistent with zero (0.008), a standard deviation, $\sigma
= 0.034$, and clear outliers make up less than 2 per cent of the
total. A comparison of $z_{\rm spec}$ versus $z_{\rm phot}$ can be
found in fig. 2 of Cirasuolo et al.\ (2008). This accuracy is
comparable to the best available from other surveys such as GOODS and
COSMOS (Caputi et al.\ 2006a; Grazian et al.\ 2006; Mobasher et al.\
2007).

\subsection{Radio observations}
\label{radio-obs}

The radio data used here are described by Ivison et al.\ (2007a) and
Ibar et al.\ (2008). Briefly, we employed the National Radio Astronomy
Observatory's (NRAO\footnote{NRAO is operated by Associated
Universities Inc., under a cooperative agreement with the National
Science Foundation.}) VLA at 1,400\,MHz, with an effective bandwidth
of 4$\times$7$\times$3.125\,MHz (via correlator mode `4') and central
frequencies in the two IF pairs of 1,365 and 1,435\,MHz. This provided
a good compromise between high sensitivity (via greater bandwidth) and
the effect of bandwidth smearing (BWS), which scales with channel
width and distance from the pointing centre. We obtained data at three
positions, the vertices of an equalatoral triangle with 15-arcmin
sides, with a 9:3:1 combination of data from the VLA's A:B:C
configurations. This yielded a synthesised beamwidth of around
1.7\,arcsec {\sc fwhm}. The noise level varies across the mosaic:
165\,arcmin$^2$ at $<$7\,$\mu$Jy\,beam$^{-1}$ in Stokes {\em I},
750\,arcmin$^2$ at $<$10\,$\mu$Jy\,beam$^{-1}$ and 1345\,arcmin$^2$ at
$<$15\,$\mu$Jy\,beam$^{-1}$. A total of 563 sources were detected at
$>$5\,$\sigma$.

When imaging, we used only those data within a radius of each pointing
centre where the dimensionless BWS term, $\beta$ (fractional bandwidth
$\times$ radius, the latter in units of {\sc fwhm}), was
$<$1.225. This ensured the reduction in peak flux density due to BWS
--- a pernicious effect which can lead to the complete loss of real
sources from radio catalogues --- was always $<$37 per cent
($(1+\beta^2)^{-0.5}$). We then calculated error-weighted values of
$\beta$ at every point in the final 3-position mosaic, using these to
correct the flux densities extracted at the position of each galaxy.

\section{Stacking the radio flux of $K$-selected populations}

\subsection{IR-selection criteria}
\label{IRsel}

\begin{figure}
\includegraphics[width=9.2cm]{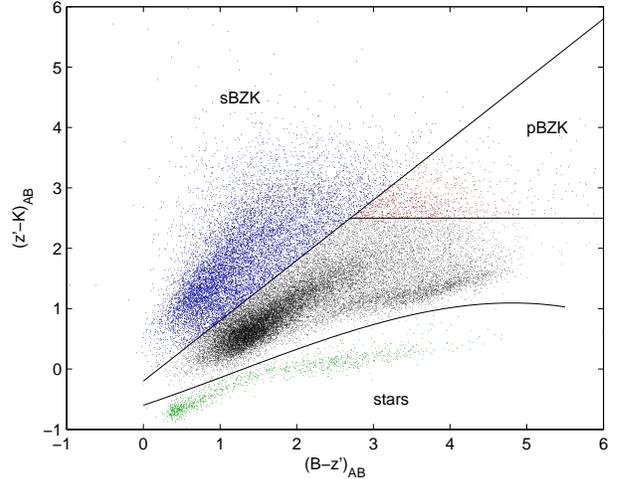}
\caption{\label{BzKF} BzK colour-colour diagram showing the location
of the different samples. The stellar branch was removed before
stacking.}
\end{figure}

The selection criteria for the galaxies to be stacked were $K \leq
23.0$ and total $\chi^2 < 10$ for the photometric redshift. This prevents
the results becoming contaminated with unreliable sources or
redshifts. Sources had to lie in the overlap region between the UDS
$K$ data, the SuprimeCam optical images and the VLA 1,400-MHz image
(1311.1\,arcmin$^2$). The total number of $K$-selected sources in this
region, meeting the above criteria is 23,185 of which 440 have radio
counterparts at $>$5\,$\sigma$.

A BzK diagram of the sample meeting the $K$ and photometric redshift
selection criteria is shown in Fig.~\ref{BzKF}. This splits into
several sub-samples based on the BzK colour cut (Daddi et al.\ 2004),
where BzK $\equiv (z^{\prime}-K) - (B-z^{\prime})$:
 
\begin{enumerate}
\item{\bf All galaxies}: all objects classified as galaxies (and not
lying on the stellar branch of the BzK diagram) meeting the selection
criteria. All points on Fig.~\ref{BzKF}, except the green stars.
\item{\bf Star-forming, high-redshift}: using the sBzK selection of
$BzK \geq -0.2$ we select a sample of 6,626 high-redshift SFGs. These
are the blue points on Fig.~\ref{BzKF}.
\item{\bf Passive, high-redshift}: using the pBzK selections of $BzK <
-0.2$ and $(z^{\prime} - K) > 2.5$ we select a sample of 542
high-redshift passive galaxies. These are the red points on
Fig.~\ref{BzKF}.
\item{\bf Non-BzK galaxies}: all of the black points in
Fig.~\ref{BzKF}. Generally galaxies at lower redshifts than the BzK
sources.
\item{\bf EROs}: defined as all objects with $(R-K)_{\rm AB} >
  3.6$. Generally these span the space overlapping pBzK, sBzK and
  non-BzK at the top right of the BzK diagram (see Fig.~\ref{erobzk}).
\end{enumerate}

Note that a sparse stellar branch is visible on the plot because the
BzK diagram is efficient at selecting stars up to 1\,mag deeper than
the compactness selection used to make the $K$ catalogue.

These colour selections are untested at the depths of the UDS and we
will investigate their reliability as best we can using the radio
data and photometric redshifts. We also note that the sBzK
star-forming sample is known to contain a fraction of AGN (Daddi et
al.\ 2004; Reddy et al.\ 2005) and we will discuss this in more
detail in \S\ref{AGN}.

\subsection{\label{stack}Radio Stacking}

\begin{figure*}
\includegraphics[width=6cm]{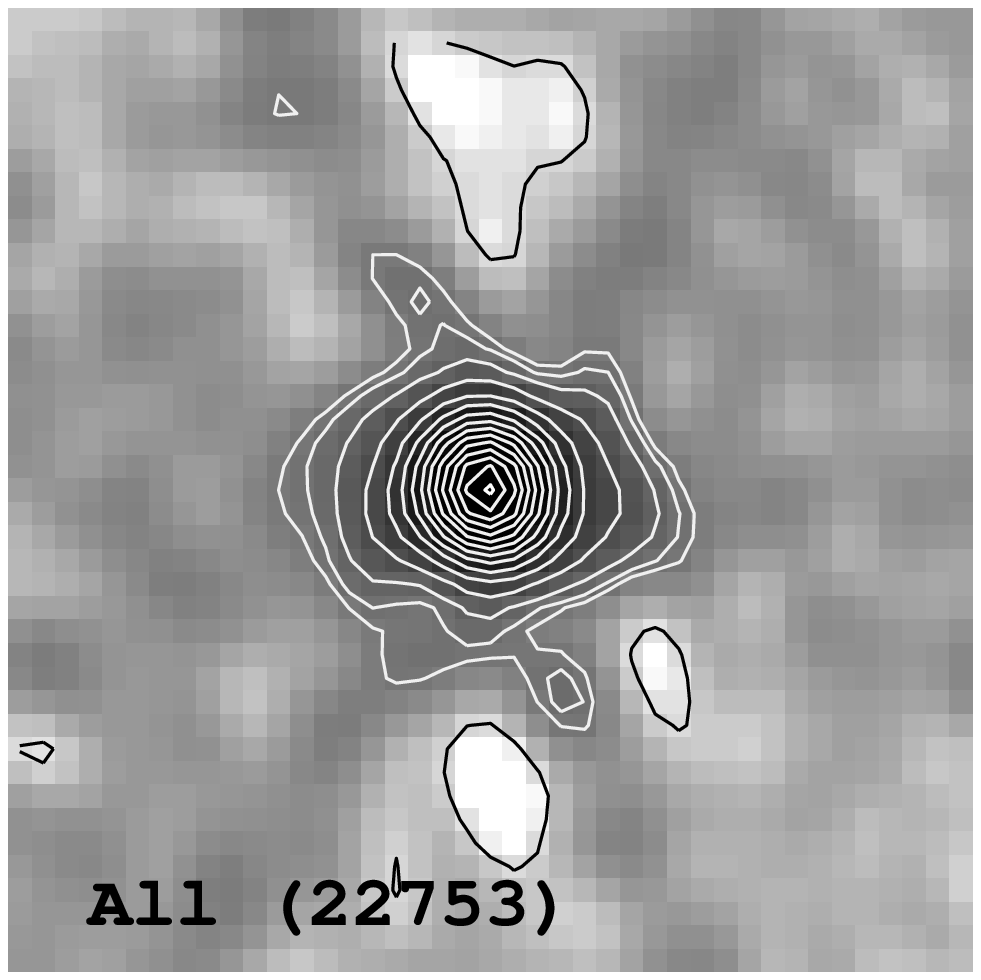}
\hspace{1cm}
\includegraphics[width=6.35cm]{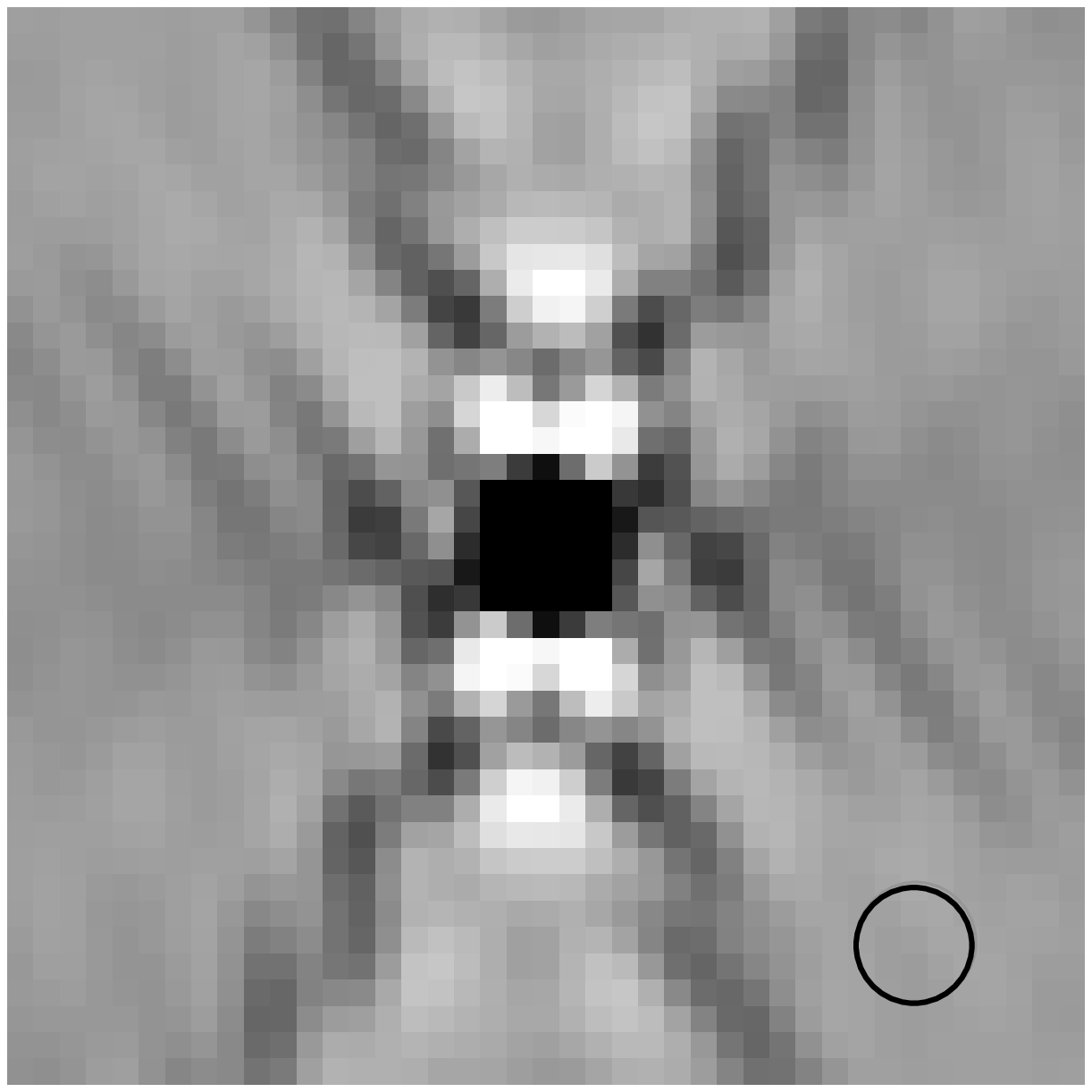}
\caption{\label{Kall_postage} {\em Left:\/} $16\times 16$-arcsec$^2$
postage-stamp image of all $K$-selected galaxies meeting the selection
criteria. The number of galaxies in the stack is indicated on the
image. Contours are at $-3,3,4,6,10,15$ and then in steps of
$+5\sigma$. S/N is so high that the lobes of the dirty beam can be
seen. {\em Right:\/} A dirty beam, averaged over the three VLA
pointings, is shown alongside at the same scale for comparison. The
size of the VLA restoring beam ($\rm FWHM = 1.7$-arcsec) is shown in
the bottom right corner.}
\end{figure*}

To perform the stacking we measured the noise-weighted mean and median
of the pixels on the radio image corresponding to the positions of the
$K$-band sources. As the radio image has units of Jy\,beam$^{-1}$,
this pixel stacking gives the total flux of a source which is no
larger than the beam. We only considered regions on the radio image
with $\sigma_{\rm 1,400} \leq 20\,\mu$Jy. For the mean stacks we
excluded pixels with $\rm|S/N|\geq 5$ but for the median we included
all sources. The errors on the median were calculated using the
distribution of flux densities, following Gott et al.\ (2001). While
there is often a small difference between the mean and median value,
these are usually within the $\sigma_{\rm 1,400}$ uncertainties in
each case. The median (Gott et al.\ 2001) is a more robust measure of
a population which may deviate from a normal distribution (as a
population containing detections will do) and so, in general, we
prefer the median. This is important in the samples which we later bin
by stellar mass or absolute $K$ mag, since as many as 30 per cent of
the sources may be detected at $>$5\,$\sigma$ in the highest bins of
stellar mass. To discount these detections (as done when computing the
mean) would not produce a representative average for this high-mass
sub-sample; with the median we can use the same statistic whatever the
fraction of detections.

We corrected the final mean or median values for the effect of
bandwidth smearing (as described in \S\ref{radio-obs}) by
multiplying the value by the mean or median BWS term. The mean and
median BWS term was calculated in the same way as the mean and median
flux density. The reason the correction was applied to the final
stacked value rather than the individual pixel values is explained
next. The BWS correction should not affect the noise on the map; it
merely corrects for the reduction in peak flux density due to
BWS. However, in stacking (where we expect no detections) our stacked
values can be positive or negative, thus the effect of a BWS
correction applied to each pixel would be to increase the effective
$\sigma_{\rm 1,400}$ on the resulting pixel flux distribution. If
we multiply each pixel by the BWS value, but not the noise on that
pixel, then we are artificially increasing the S/N of every
pixel. This has two consequences: first, we finish with statistics
which are not Gaussian (e.g.\ more $>$3-$\sigma$ peaks on the stacked
maps than expected); second, as we are rejecting $\rm|S/N| \geq
5$ points from the stack, we have fewer points to stack. If we decide
to increase the noise at every point by the BWS factor to counter
these problems then, again, we have less sources to stack as a larger
fraction of the map then appears to have $\sigma_{\rm 1,400} >
20\,\mu$Jy. Our solution was to multiply both the final flux density
in the stack {\em and\/} the final error by the averaged BWS. This
leaves us with the correct number of sources and Gaussian statistics
(so the reliability of the detections is intuitive to calculate).

\begin{table*}
\caption{\label{aipsfit} Integrated fluxes and fitted sizes of postage
stamps.}
\begin{tabular}{ccccccc}
\hline
\multicolumn{1}{c}{Redshift}&\multicolumn{1}{c}{$K$ cut}&\multicolumn{1}{c}{\Smean}&\multicolumn{1}{c}{$S_{\rm 1,400}^{\rm med}$}&\multicolumn{1}{c}{$S_{\rm 1,400}^{\rm int}$}&\multicolumn{1}{c}{Ratio}&\multicolumn{1}{c}{Size}\\
\multicolumn{2}{c}{}&\multicolumn{1}{c}{($\mu$Jy)}&\multicolumn{1}{c}{($\mu$Jy)}&\multicolumn{1}{c}{($\mu$Jy)}&\multicolumn{1}{c}{}&\multicolumn{1}{c}{(arcsec)}\\
\hline
$z<0.2$ & $K\leq 23.0$ & $3.34\pm 0.42$ & $2.99\pm0.74$ &
$7.64\pm1.59$ & $2.29\pm 0.56$ & $3.96\,(0.66) \times 2.66 \,(0.44)$\\
$z<0.7$ & $K\leq 23.0$ & $4.82\pm 0.15$ & $4.67\pm 0.23$ & $7.04\pm
0.43$ & $1.46\pm 0.10$ & $2.52\,(0.11) \times 2.34\,(0.10)$\\
$z>0.7$ & $K\leq 23.0$ & $6.44\pm 0.11$ & $6.15\pm 0.19$ & $ 8.24\pm
0.35$ & $1.28\pm 0.06$ & $2.32\,(0.07)\times 2.17\,(0.06)$\\
$z>0.2$ & $K\leq 21.5$ & $10.13\pm 0.16$ & $10.44\pm0.28$ &
$14.15\pm0.56$ & $1.40\pm 0.06$ & $2.43\,(0.07)\times 2.28\,(0.06)$\\
$z>0.2$ & $21.5 < K \leq 23.0$ & $4.15\pm 0.11$ & $3.90\pm 0.17$ &
$5.07\pm 0.27$ & $1.22\pm 0.07$ & $2.29\,(0.08)\times 2.09\,(0.07)$\\
\hline
\end{tabular}  
\flushleft
\footnotesize{Columns (1) and (2): redshift and $K$ cuts for the
postage-stamp images; (3) noise-weighted mean flux, excluding
$\geq$5-$\sigma$ points (corrected for BWS); (4) median flux and
68-per-cent confidence interval (CI) from pixel stack (BWS-corrected);
(5) integrated flux and error from {\sc jmfit}; (6) ratio between
integrated and mean flux; (7) fitted size in {\sc jmfit}, with
uncertainties on the major- and minor-axis {\sc fwhm} in parentheses.}
\end{table*}

We performed exactly the same analysis at random positions to check
that there were no systematic effects in the radio map which might
lead us to a positive flux. Flux was stacked at 21,000 random
positions and this stack was repeated 1,000 times. The mean of the
1,000 mean values was 0.015\,$\mu$Jy with $\sigma_{\rm
1,400}=0.13\,\mu$Jy and the mean of the 1,000 medians was
$-0.009\,\mu$Jy with $\sigma_{\rm 1,400}=0.17\,\mu$Jy. This is
consistent with zero and well below any fluxes quoted from a stack of
$K$-selected objects.

We also took a $41\times 41$-pixel region around every IR position and
stacked this to produce a noise-weighted mean image stack (see
Fig.~\ref{Kall_postage}). We excluded any pixels on the radio image
with a high noise level ($>$20\,$\mu$Jy) or with a value of
$\rm|S/N|\geq 5$). These postage-stamp images are not corrected for
BWS as the integrated flux is not affected by band-width smearing. The
integrated flux is measured directly from the postage stamp images
using {\sc jmfit} in \AIPS and also includes any systematic effects
due to positional uncertainty in the $K$-band positions, as well as
any flux from sources which are larger than the beam. We made stacks
of samples cut in redshift and $K$ to see if either produced a more
significant difference to the pixel stacks, as we might expect
positional uncertainty to be greater at fainter $K$ and source sizes
to be larger at lower redshift. The results are listed in
Table~\ref{aipsfit}. We found that samples which were cut at $z>0.2$
gave consistent ratios between the \AIPS\ integrated fit and the mean
from the pixel stack. Thus positional errors on the $K$-selected
sources did not vary significantly as a function of $K$ mag to our
limit of $K= 23.0$. The largest effect seems to be from partially
resolved sources at low redshifts. The average ratio of integrated to
BWS-corrected mean-weighted-stacked flux density is 1.31 for $z>0.2$
and 2.29 for $z<0.2$. A further correction needs to be made to this
integrated flux for the fact that these radio stacks are not `{\sc
clean}ed'. In the VLA image, only sources visible to the eye were {\sc
clean}ed, via clean boxes placed by hand. Sources at the low flux
density levels found here resemble the dirty beam with flux (both
positive and negative) in their sidelobes (see
Fig.~\ref{Kall_postage}). We looked at the differences between a
source fitted in a cleaned map compared to a dirty map and found the
average factor between the {\sc clean}-fitted flux to dirty-fitted
flux to be 1.14. Thus the total correction we need to make to our
fluxes is $1.14\times 1.31 = 1.49$ for $z>0.2$ and $1.14\times 2.29 =
2.61$ for $z<0.2$. Fluxes labelled \Sint\ have been corrected in this
way. All derived quantities in figures and tables ($L_{\rm 1,400}$,
SFR, SSFR) have been corrected.

We also checked that the noise integrated down as expected following
Poissonian statistics by measuring the noise in the stacked
postage-stamp images around the detected sources and on the
Monte-Carlo images. This confirmed that the noise does integrate down
as expected, i.e.\ as $\sigma/\sqrt{N_{\rm sources}}$.

\subsection{Deriving SFRs and luminosities}

We can use photometric redshifts at the time of stacking the radio
data to derive rest-frame 1,400-MHz luminosities, $L_{\rm 1,400}$,
and SFRs for the galaxies.  The 1,400-MHz luminosity is calculated for
each pixel based on the flux at that pixel and the redshift of the
galaxy being stacked, using

\[
L_{\rm 1,400} \,({\rm W\,Hz^{-1}}) = 9.52\times 10^{18}\, S_{\rm 1,400}\,
D_{\rm L}^2 \, 4\pi (1+z)^{-0.2}
\]

\noindent
where $D_{\rm L}$ is the luminosity distance of an individual galaxy
in Mpc and $S_{\rm 1,400}$ is the 1,400-MHz flux density of the pixel
corresponding to that galaxy in Jy. This assumes a radio spectral
index ($S_{\nu} \propto \nu^{\alpha}$) of $\alpha = -0.8$.

The median 1,400-MHz luminosity is then the median of the luminosities
for the individual pixels. We note that this is not the same as taking
the median stacked flux and median redshift and applying the above
equation, since the median redshift of the sub-sample may not be that
at which the peak of the radio emission is being produced.

The SFR is then calculated from the median 1,400-MHz luminosity in two
ways. First, following Condon (1992), Haarsma et al.\ (2000) and
Condon et al.\ (2002), we have

\[
\dot{\rho_{\ast}} ({\rm M_{\odot} \,yr^{-1}}) = \left[ \frac{5.5 \times
L_{\rm 1,400}}{(5.3\times 10^{21}\times \nu^{-0.8}) + (5.5\times 10^{20}
\times \nu^{-0.1})}\right] 
\]

\noindent
where $\nu$ is the frequency in GHz. For 1,400\,MHz, this becomes

\[\dot{\rho_{\ast}}({\rm M_{\odot} \,yr^{-1}}) = 1.2006\times 10^{-21}
L_{\rm 1,400}\]

Second, following Bell (2003), we have

\begin{eqnarray*}
\dot{\rho_{\ast}} ({\rm M_{\odot} \,yr^{-1}}) & = & 5.52\times 10^{-22}\, L_{\rm 1,400},\,\,\,\,\, L
> L_c  \\
  & = & \frac{5.52\times 10^{-22}\, L_{\rm 1,400}}{0.1 + 0.9(L/L_c)^{0.3}} ,\,\,\,\,
L\leq L_c \\
\end{eqnarray*}
where $L_c = 6.4\times 10^{21}\,\rm{W\,Hz^{-1}}$.

Both conversions assume a Salpeter-like initial mass function (IMF)
with $\Psi(M) \propto M^{-2.35}$ and a mass range of $0.1-100
\,\rm{M_{\odot}}$. The study by Bell (2003) suggests that the radio
does not trace star formation as effectively at low values of $L_{\rm
1,400}$ because the non-thermal radio emission is suppressed by a
factor of 2--3 in $\sim L_{\ast}/100$ galaxies relative to
$L_{\ast}$. The Bell (2003) conversion thus boosts the radio SFR at
low $L_{\rm 1,400}$ values by a factor $\sim 2$ compared to Condon
(1992); however, the overall normalisation of SFR is lower than that
used by Condon (1992).

\section{Results}
\label{res}

\begin{table*}
\caption{\label{fluxT} Stacked radio properties at 1,400-MHz for $K$-selected samples.}
\begin{tabular}{lccccccccc}
\hline
\multicolumn{1}{l}{Sample}&\multicolumn{1}{c}{N}&\multicolumn{1}{c}{\%
det}&\multicolumn{1}{c}{$\bar{S}_{\rm 1,400}$}&\multicolumn{1}{c}{$S_{\rm 1,400}^{\rm med}$}&\multicolumn{1}{c}{\Sint}&\multicolumn{1}{c}{$L_{\rm 1,400}$}&\multicolumn{1}{c}{SFR}&\multicolumn{1}{c}{SSFR}&\multicolumn{1}{c}{$\langle
z\rangle$}\\
\multicolumn{3}{c}{}&\multicolumn{1}{c}{($\mu$Jy)}&\multicolumn{1}{c}{($\mu$Jy)}&\multicolumn{1}{c}{($\mu$Jy)}&\multicolumn{1}{c}{($\rm{W\,Hz^{-1}}$)}&\multicolumn{1}{c}{($\rm{M_{\odot}\,yr^{-1}}$)}&\multicolumn{1}{c}{($\rm{Gyr^{-1}}$)}&\multicolumn{1}{c}{}\\
\hline
All galaxies & 23,185 & 1.9 & $5.90\pm 0.09$ & $5.66\pm 0.15$ & $8.43\pm
0.22$ & 
$1.34\times10^{22}$ & $16.0\pm 0.6$ & $1.137\pm 0.042$ & 0.96 \\
Non-BzK & 16,048 & 1.7 & $5.30\pm 0.11$ & $4.95\pm 0.17$ &
$7.38\pm0.25$ & $1.06\times 10^{22}$ & $8.5\pm 0.4$ &
$0.694\pm 0.031$ & 0.72 \\
sBzK & 6,626 & 2.1 & $7.42\pm 0.17$ & $ 7.47\pm 0.25$ & $11.13\pm 0.37$
& $1.28\times10^{23}$ & $153.6\pm 6.7$ & $ 5.05 \pm 0.22$ & 1.55 \\
pBzK & 542 & 3.9 & $5.97\pm 0.57$ & $ 6.72\pm 1.02$ & $ 10.01\pm 1.52$
& $1.11\times10^{23}$ & $131.0\pm 18.6$ & $2.38\pm0.39$ & 1.49 \\
EROs & 2,905 & 4.0 & $7.86\pm 0.25$ & $ 7.94\pm 0.49$ & $11.83\pm 0.73$
& $1.01\times 10^{23}$ & $121.1\pm 6.7$ & $2.49\pm 0.13$ & 1.42  \\
Stellar & 617 & 0 & $0.13\pm 0.56$ & $ 0.39\pm 0.78$ & .. & .. & .. & .. &
.. \\ 
\hline
\end{tabular}
\flushleft \footnotesize{Columns: (1) sample name; (2) number in stack
(including detections); (3) fraction detected at $\geq$5\,$\sigma$;
(4) noise-weighted mean, excluding detections; (5) median and
68-per-cent CI; (6) integrated flux (median corrected as described in
\S\ref{stack}); (7) median 1,400-MHz luminosity, corrected to
integrated value; (8) median SFR (using Condon 1992) and 68-per-cent
CI, corrected to integrated; (9) median SSFR ($=\rm{SFR/M_{\rm
stellar}}$) and 68-per-cent CI, corrected to integrated; (10) median
redshift.}
\end{table*}

\begin{figure*} 
\includegraphics[width=4.3cm]{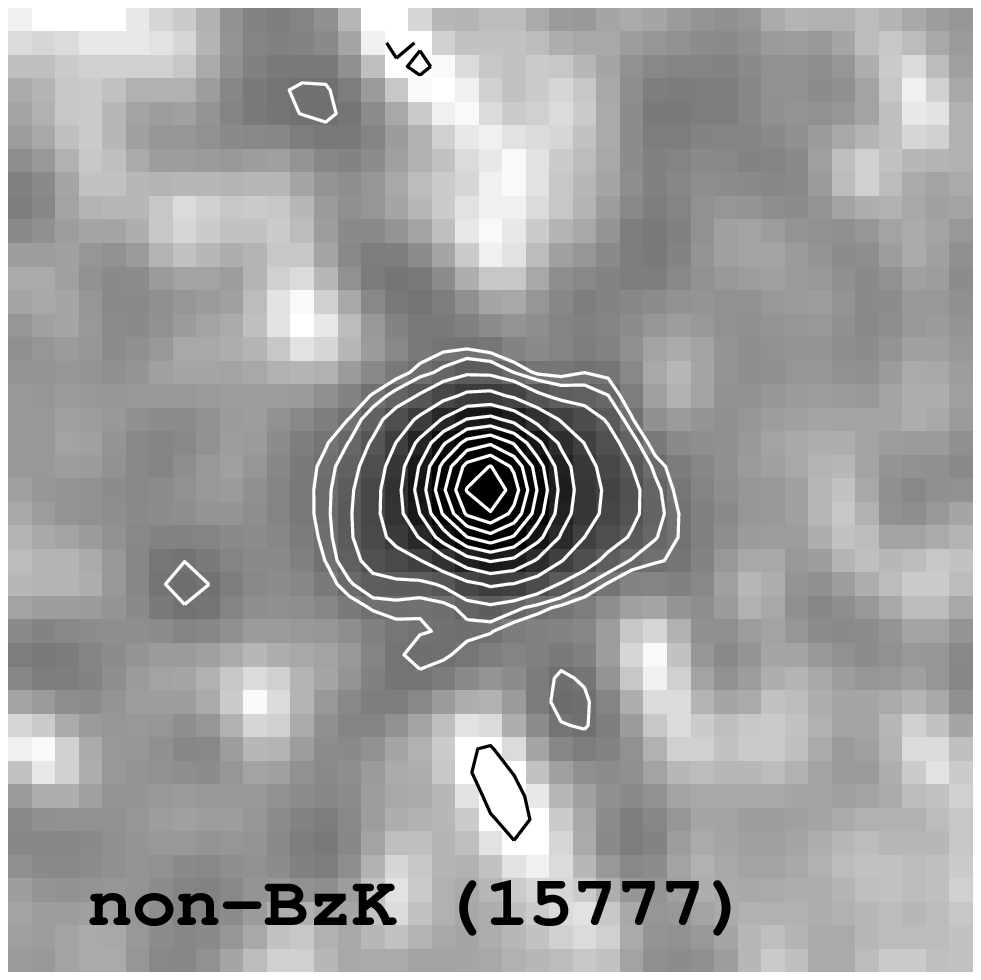}
\includegraphics[width=4.3cm]{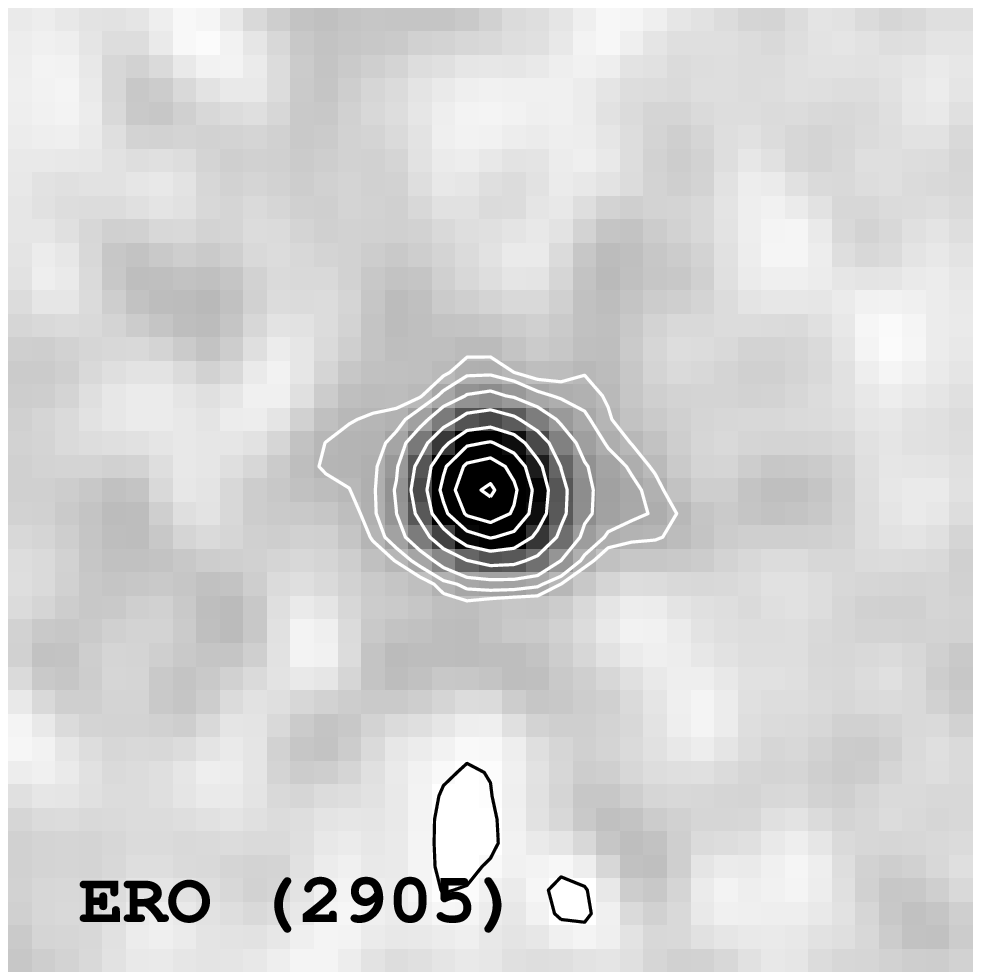}
\includegraphics[width=4.3cm]{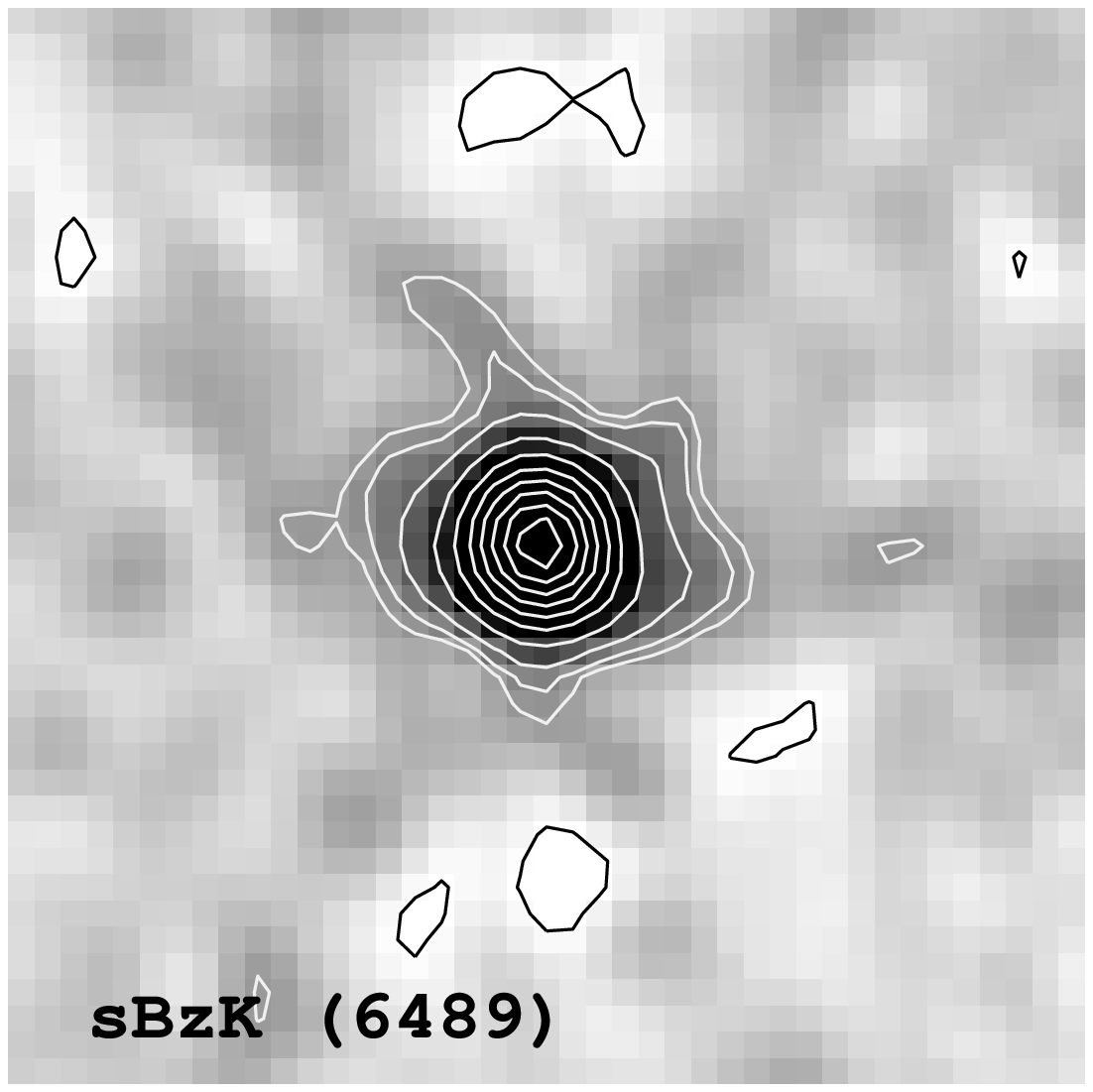}
\includegraphics[width=4.3cm]{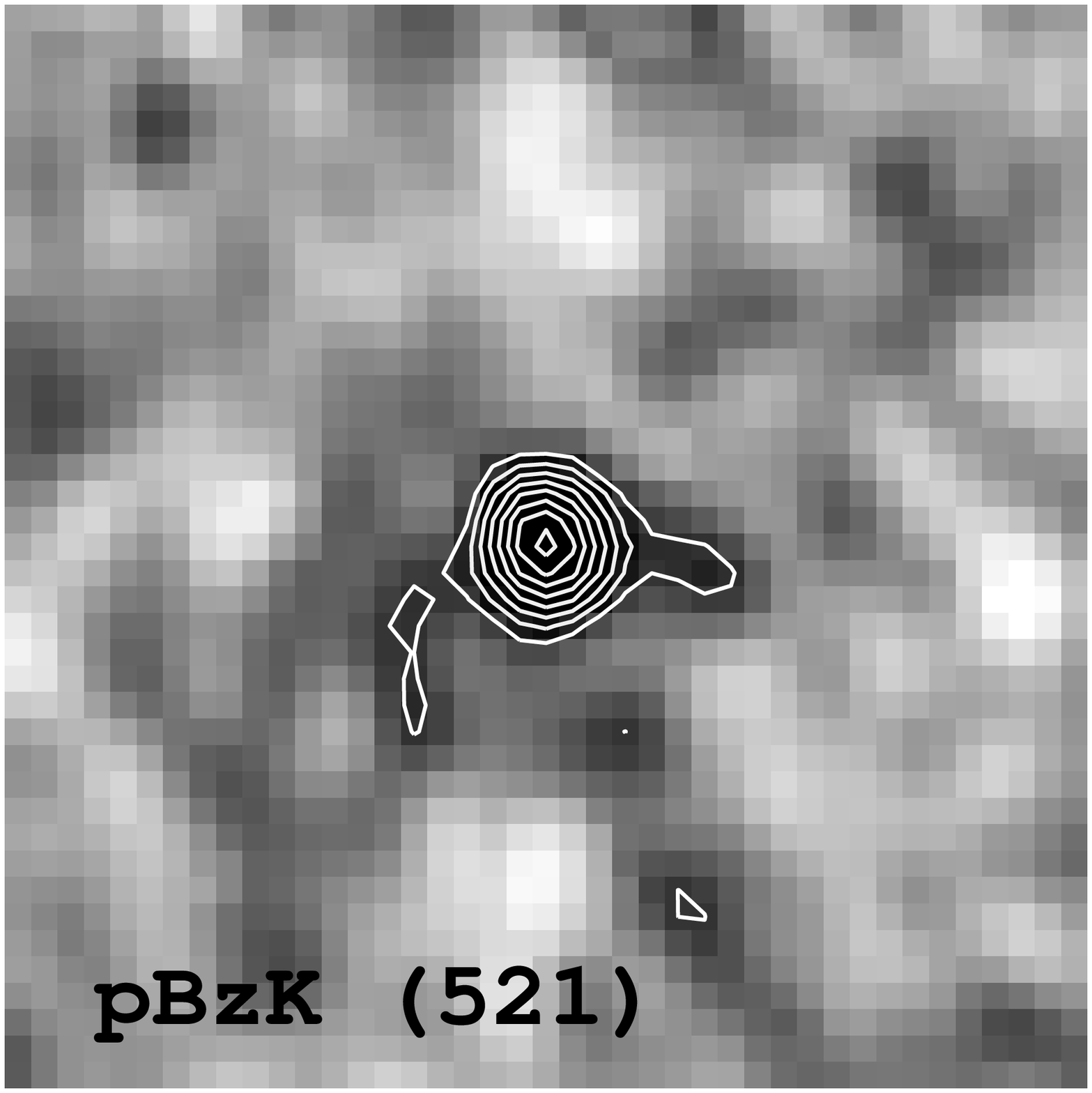}
\caption{\label{bzk_postage} Postage-stamp images of the 1,400-MHz
radio stacks for non-BzK, ERO, sBzK and pBzK galaxies in the
UDS. Numbers in parenthesis denote the number of sources in each
stack. Contours for non-BzK, ERO and sBzK galaxies are plotted at
$-3,3,4,6,10,15$ and then in steps of $+5 \sigma$, while for pBzK
galaxies they are plotted at $-3,3,4,5$ and then in steps of $+1
\sigma$. Pixels deviating by more than $\pm 5 \sigma$ from the mean
have been removed.}
\end{figure*}

\begin{center}
\begin{figure}
\includegraphics[width=7.8cm,height=4.6cm]{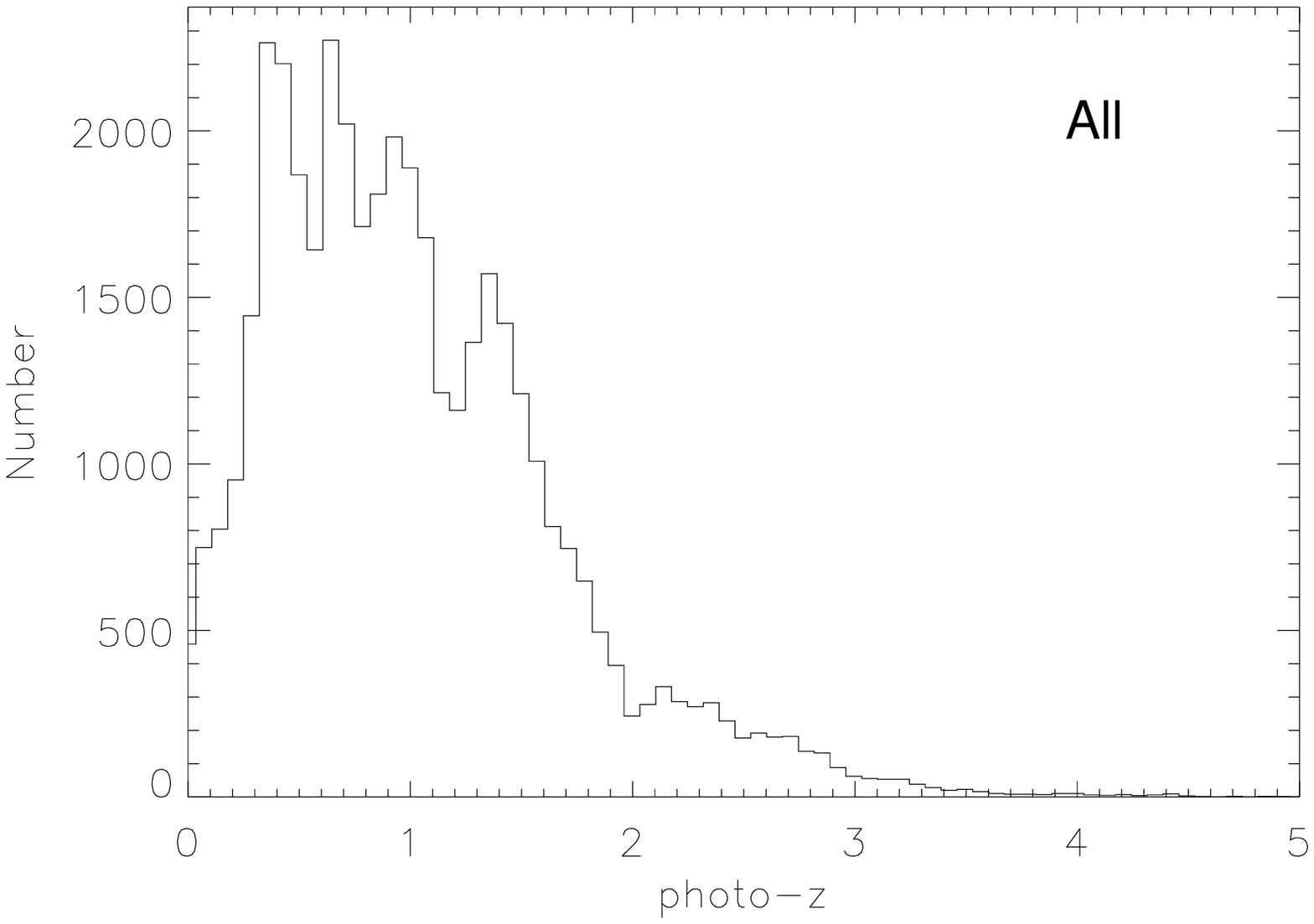}
\includegraphics[width=7.8cm,height=4.6cm]{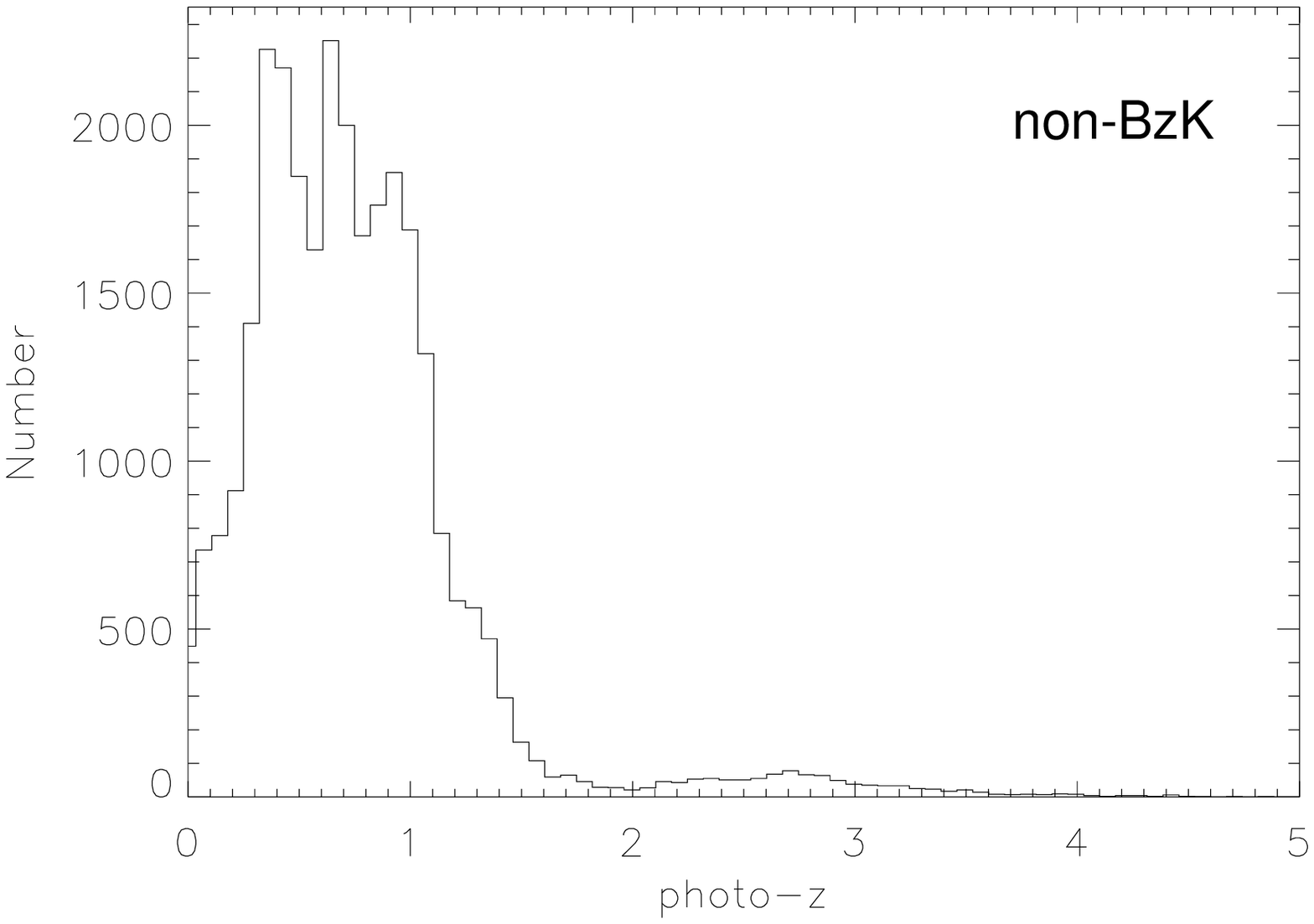}
\includegraphics[width=7.8cm,height=4.6cm]{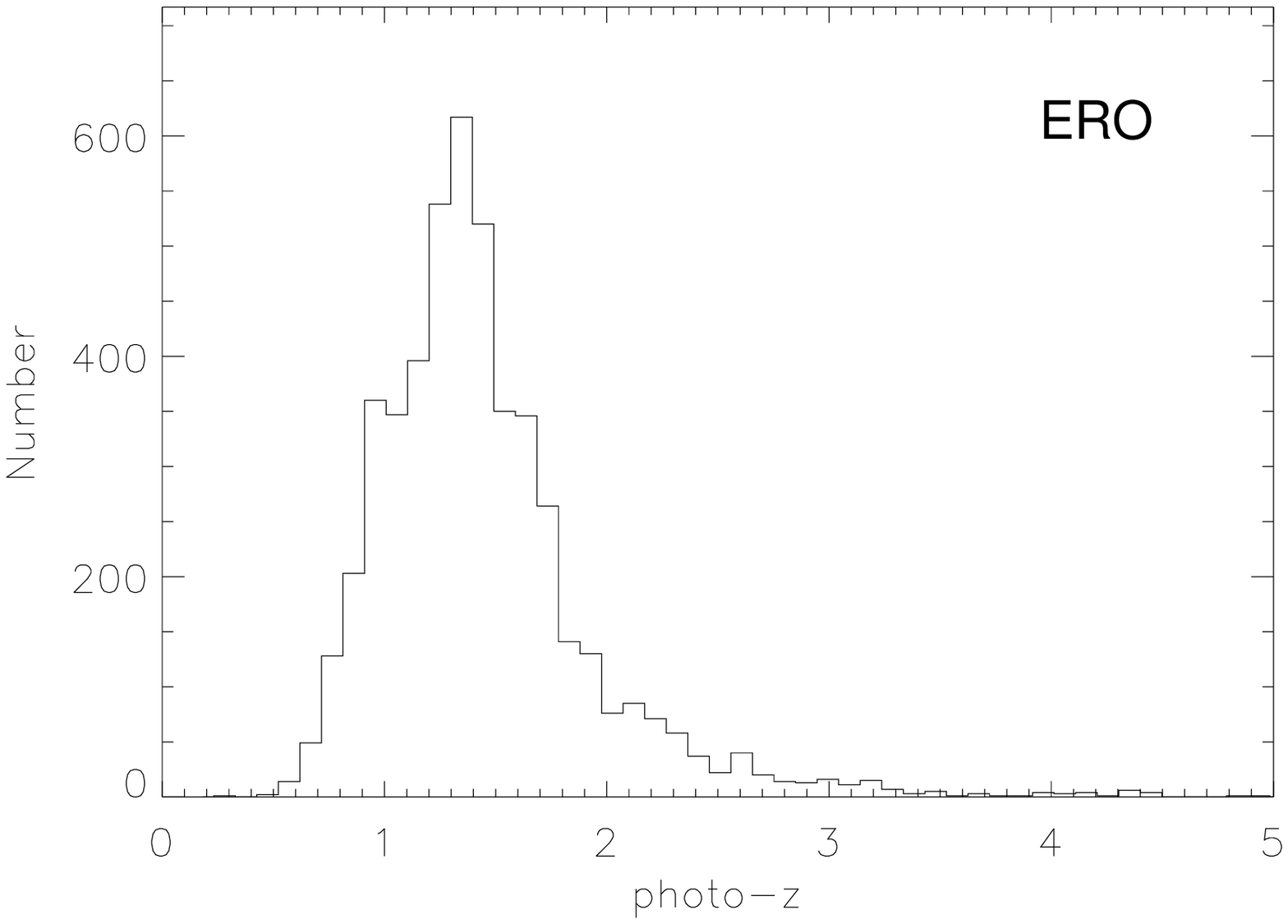}
\includegraphics[width=7.8cm,height=4.6cm]{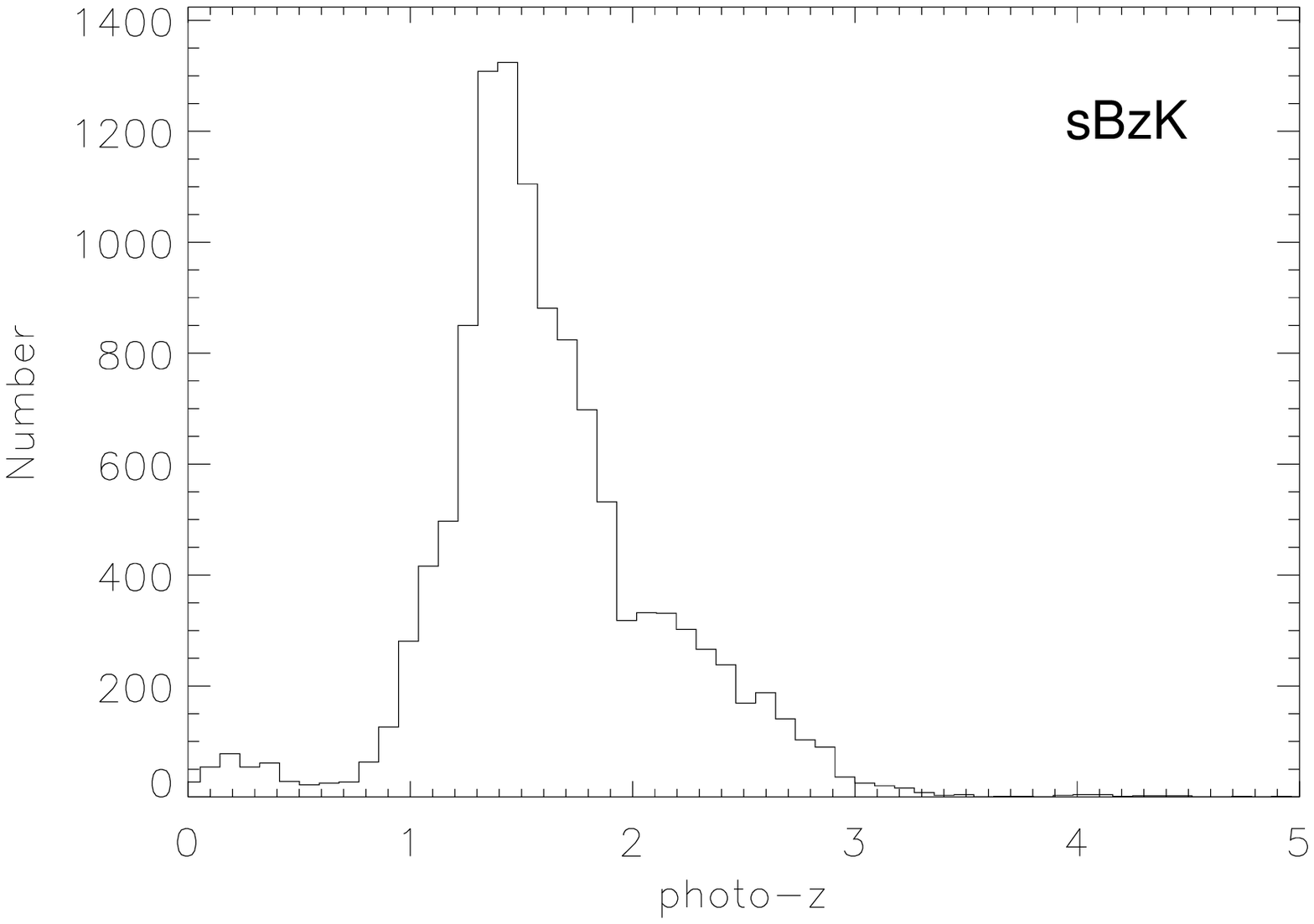}
\includegraphics[width=7.8cm,height=4.6cm]{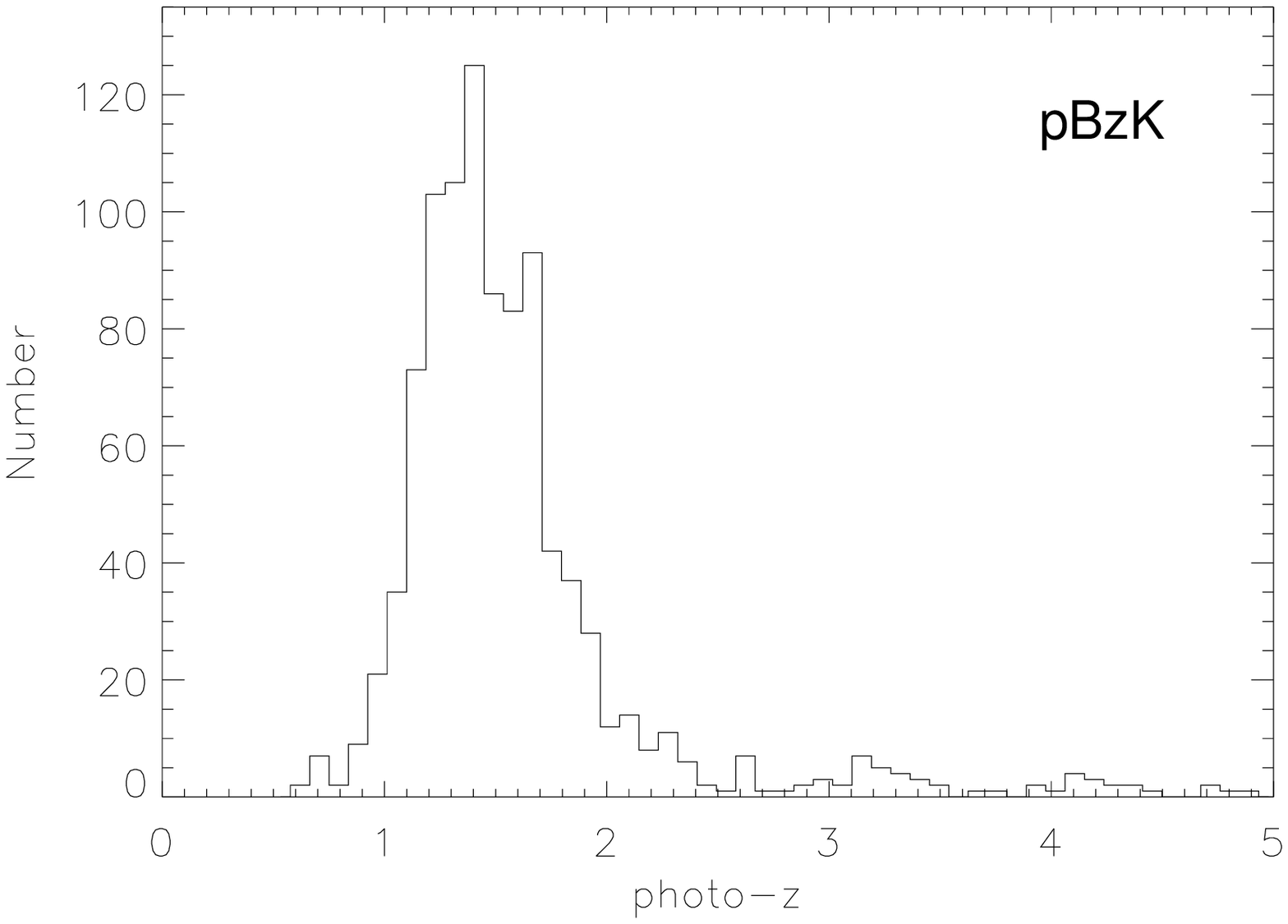}
\caption{\label{BzK_nz} Photometric redshift distributions for the
$K$-selected samples discussed in the paper.}
\end{figure}
\end{center}

Table~\ref{fluxT} shows the results for each sample and for each
stacking method. All four sub-sets of $K$-selected galaxies (ERO,
non-BzK, sBzK and pBzK) are significantly detected in the stacks,
their postage-stamp stacked images are shown in Fig.~\ref{bzk_postage}.

The BzK selection (Daddi et al.\ 2004) is designed to select either
star-forming (sBzK) or passively-evolving (pBzK) galaxies in the
redshift range $1.4<z<2.5$. We have taken the colour cut used by Lane
et al.\ (2007), as shown in Fig.~\ref{BzKF}, which gives us samples of
6,626 sBzK and 542 pBzK galaxies that match the selection criteria and
overlap with our radio data. This is by far the largest sample of such
galaxies to be stacked in the radio to date.

The redshift distributions for the various samples are shown in
Fig.~\ref{BzK_nz}. It is immediately apparent that the distributions
for BzK galaxies have tails at redshifts higher and lower than the
range in which they are designed to lie. At $z<1.4$, this amounts to a
fraction of 37 (29) per cent for pBzK (sBzK) galaxies. There are, of
course, uncertainties in the photometry which may shift objects in and
out of the BzK regions and, for pBzK, the distribution of colours
means that more galaxies are likely to scatter in rather than out,
accounting for the raised fraction compared to sBzK galaxies. There
may also be errors in the photometric redshifts which place some
galaxies at a spuriously low redshift. For sBzK galaxies, the $z<1.4$
objects mostly lie along the boundary with non-BzK objects, suggesting
that photometric scatter is the main cause of the low-redshift
tail. For pBzK galaxies, however, the $z<1.4$ objects fill the same
region of colour space as those at $z>1.4$. We will investigate the
radio properties of BzK galaxies at all redshifts to see if there are
any differences in the $z<1.4$ populations which are contaminating the
BzK space. We note that a spectroscopic study by Popesso et al.\
(2008) also finds that 23 percent of sBzK lie at $z<1.4$. The number
of pBzK in their sample is too small to produce a reliable measure of
the fraction at low redshift.

\begin{figure*}
\includegraphics[width=8.8cm]{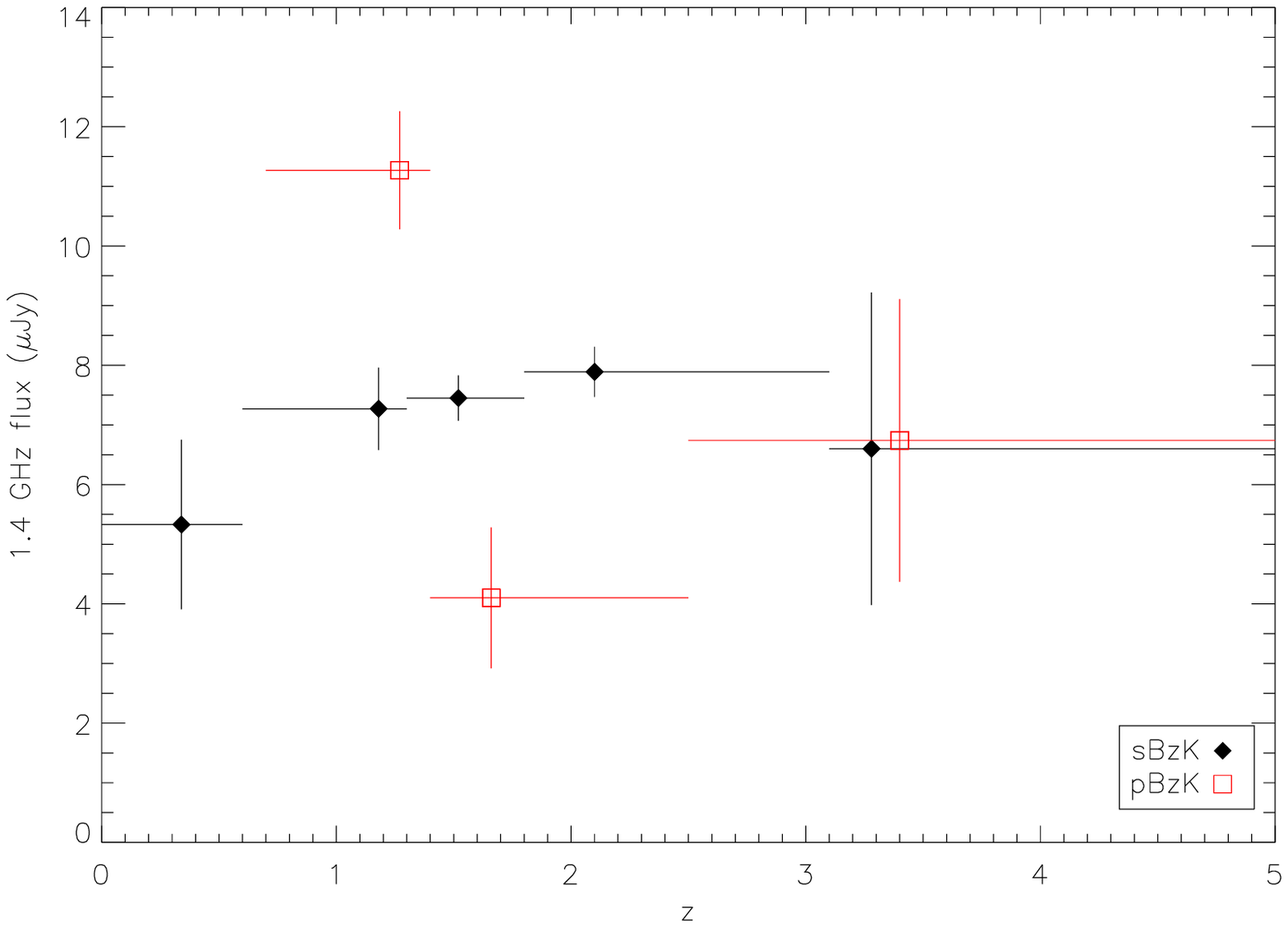}
\includegraphics[width=8.8cm]{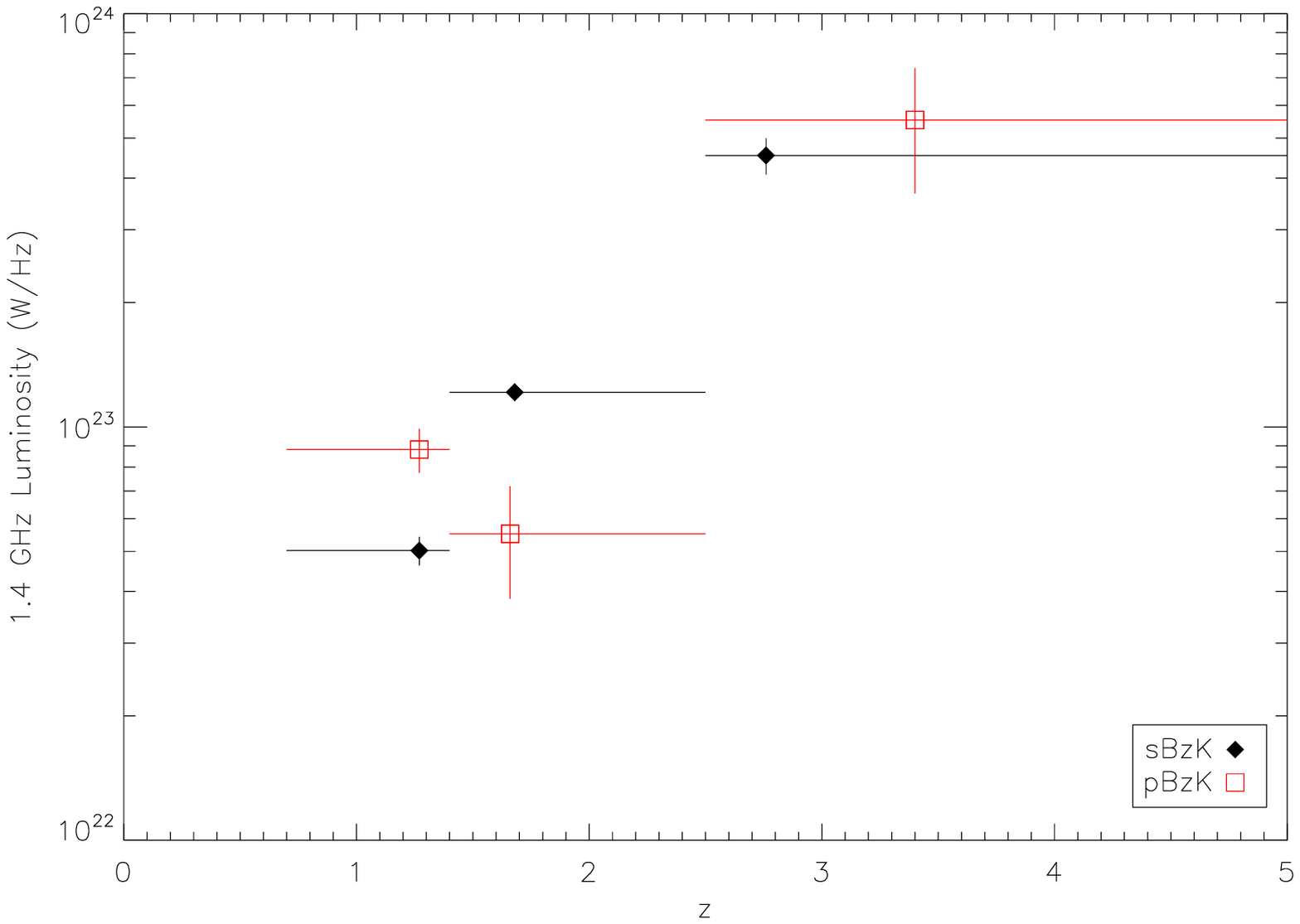}
\caption{\label{fluxzF} {\em Left}: Stacked median flux of BzK galaxies as a
function of redshift. Solid black diamonds are sBzK galaxies and red
open squares are pBzK galaxies. The sBzK galaxies show a roughly
constant flux with increasing redshift while the pBzK galaxies suffer
a dramatic reduction in radio flux in the $1.4<z<2.5$ range. This plot
argues that the $z<1.4$ pBzK galaxies are not merely redshift outliers
but are a physically distinct population. {\em Right}: Stacked median
luminosity of BzK galaxies as a function of redshift, same symbols as
before. The luminosity of pBzK galaxies drops over the interval
$0.7<z<2.5$ whereas for sBzK it is rising strongly.}.
\end{figure*}

\begin{figure}
\includegraphics[width=8.5cm]{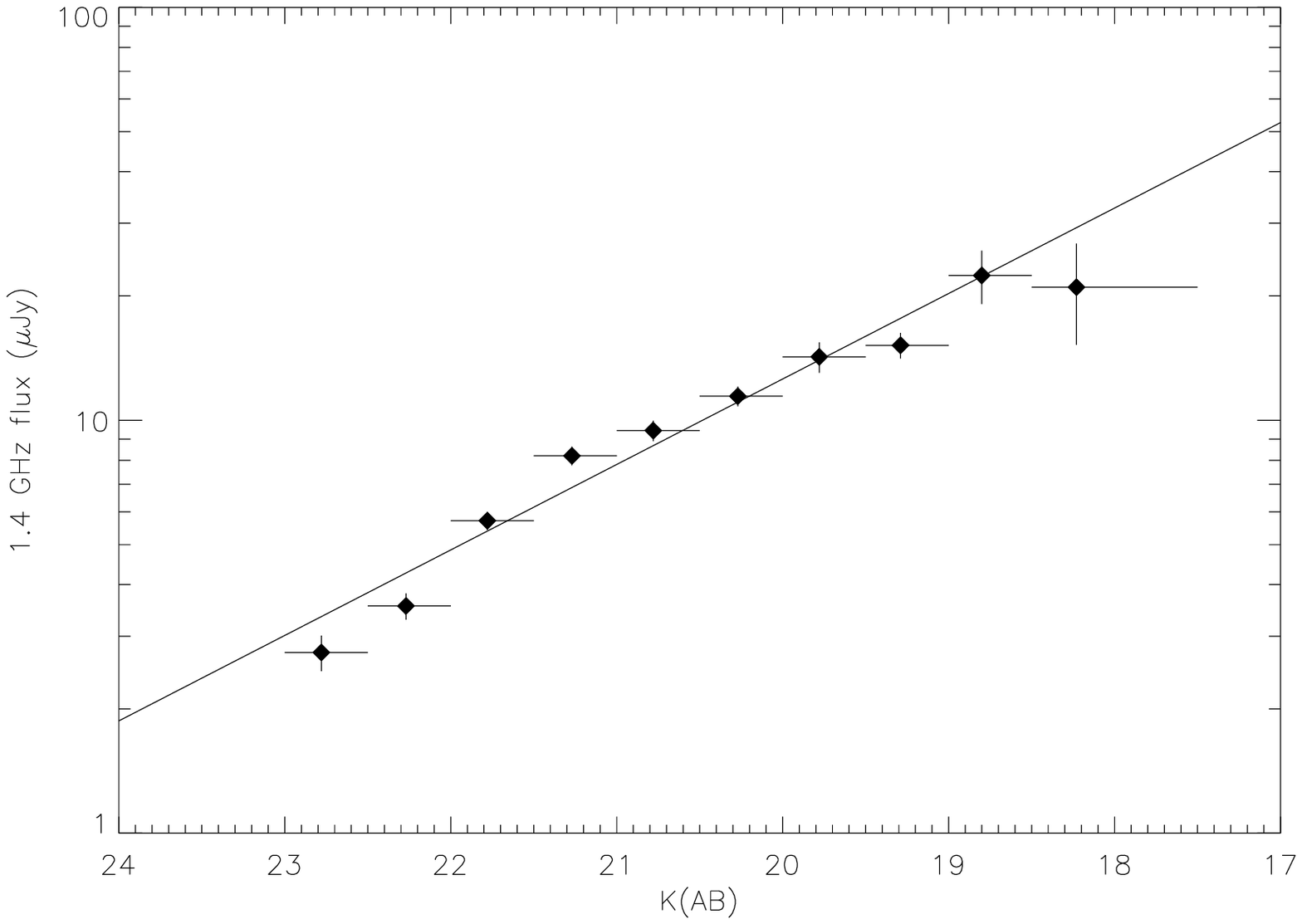}
\includegraphics[width=8.5cm]{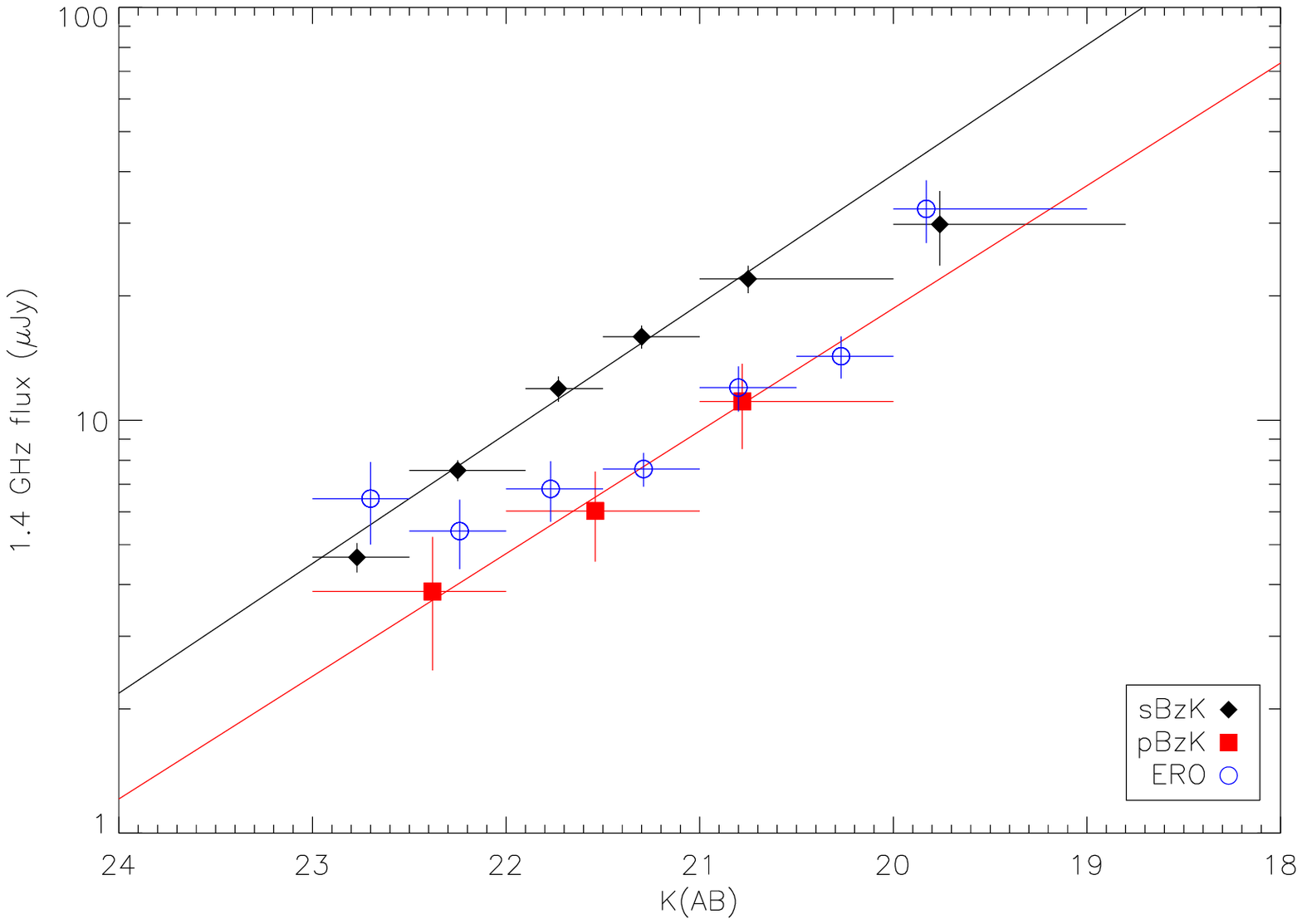}
\includegraphics[width=8.5cm]{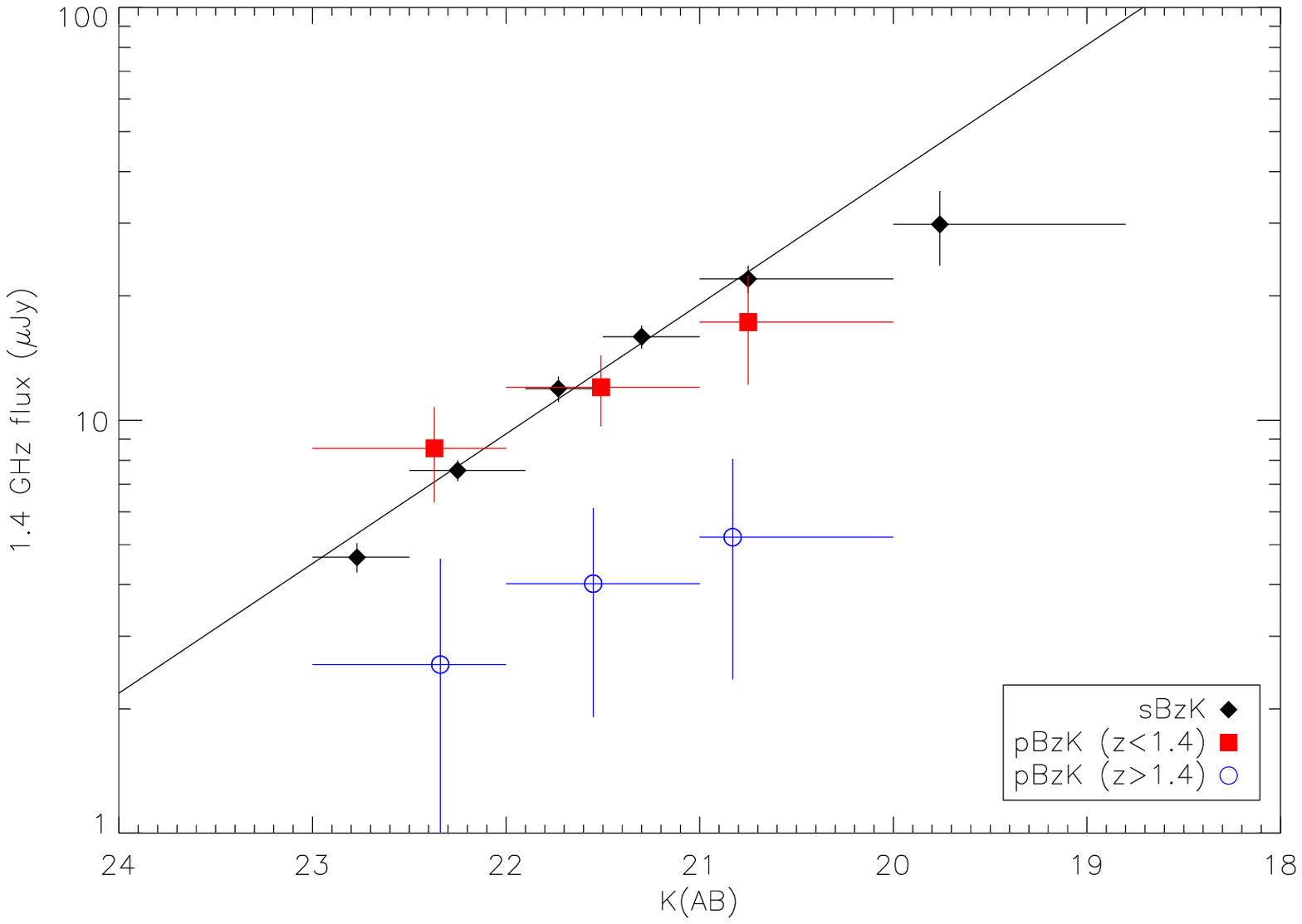}
\caption{\label{fluxKF} Relationship between $K$ and median
stacked 1,400-MHz flux density for all galaxies {\em (top)} and BzK
and ERO galaxies {\em (middle)}. Best linear fits are plotted. There
is no power-law fit for EROs. Both sBzK and pBzK galaxies have a
similar slope with $S_{\rm 1,400}\propto S_K^{0.74-0.79}$, though the
pBzK sample shows less radio flux density at a given $K$ mag. EROs
seem to follow a similar trend to pBzK galaxies at $K > 21.5$, while
at fainter fluxes the correlation with $K$ seemingly breaks down. In
the full sample and the ERO sample, the median redshift increases as
$K$ mag increases. {\em Bottom:} pBzK galaxies are split into two redshift
bins, $z<1.4$ and $1.4<z<2.5$ matching the transition seen in
Fig.~\ref{fluxzF}. The low-redshift pBzK galaxies are seen to follow the same
radio-$K$ relation as the sBzK galaxies, supporting the idea that they are
star-forming interlopers. The pBzK galaxies at $1.4<z<2.5$ are not
detected significantly in the radio and have a radio-$K$ relation offset
from that of sBzK and low-redshift pBzK galaxies.}
\end{figure}

In Fig.~\ref{fluxzF} (left) we show the median flux of BzK galaxies as
a function of redshift. The average 1,400-MHz flux density of sBzK
galaxies is roughly constant (or slowly rising) over the interval
$0.6<z<3.0$, implying a strong evolution in luminosity. For pBzK
galaxies there is a dramatic change in the average radio emission from
$0.7<z<1.4$ to $1.4<z<2.5$. At $z<1.4$, the pBzK galaxies are
relatively bright radio emitters, and are in fact more luminous at
this redshift than sBzK galaxies. At $z>1.4$, their radio emission
falls to approximately half that of sBzK galaxies. This is the first
piece of evidence that the pBzK sample (as defined by the BzK diagram)
is not a homogeneous population of passive galaxies. The low-redshift
tail cannot be due to photometric errors as the radio properties
should then be consistent across the redshift boundary, which they
clearly are not. In Fig.~\ref{fluxzF} (right) we also plot median
\Lrad\ against redshift for the sBzK and pBzK samples. The flat
flux-density-redshift trend for sBzK galaxies in Fig.~\ref{fluxzF}
(left) translates into a steep rise of \Lrad\ with redshift. In
contrast, the pBzK sample behaves quite differently, with median
\Lrad\ falling slightly over the range $0.7<z<2.5$ before rising
dramatically at higher redshifts (in line with sBzK galaxies). The
radio luminosity of pBzK galaxies is higher than that of sBzK galaxies
at $0.7<z<1.4$; however, the pBzK galaxies are also more luminous in
$K$ over this redshift interval. The elevated luminosity of the pBzK
galaxies in this redshift bin is due to the correlation between radio
and $K$ fluxes and luminosities (see Fig.~\ref{fluxKF} and
~\ref{sfrKabsF}). Indeed, the ratio of radio to $K$ fluxes in sBzK and
pBzK galaxies at $0.7<z<1.4$ is the same (see Fig.~\ref{fluxKF}).
pBzK galaxies are also more luminous in $K$ than sBzK galaxies in the
$1.4<z<2.5$ bin. This means that the difference between the sBzK and
pBzK galaxies in this bin in Fig.~\ref{fluxzF} (right) is
intrinsically larger when normalised to $K$ luminosity than is shown
here.

\begin{figure*}
\includegraphics[width=4.7cm,height=4.5cm]{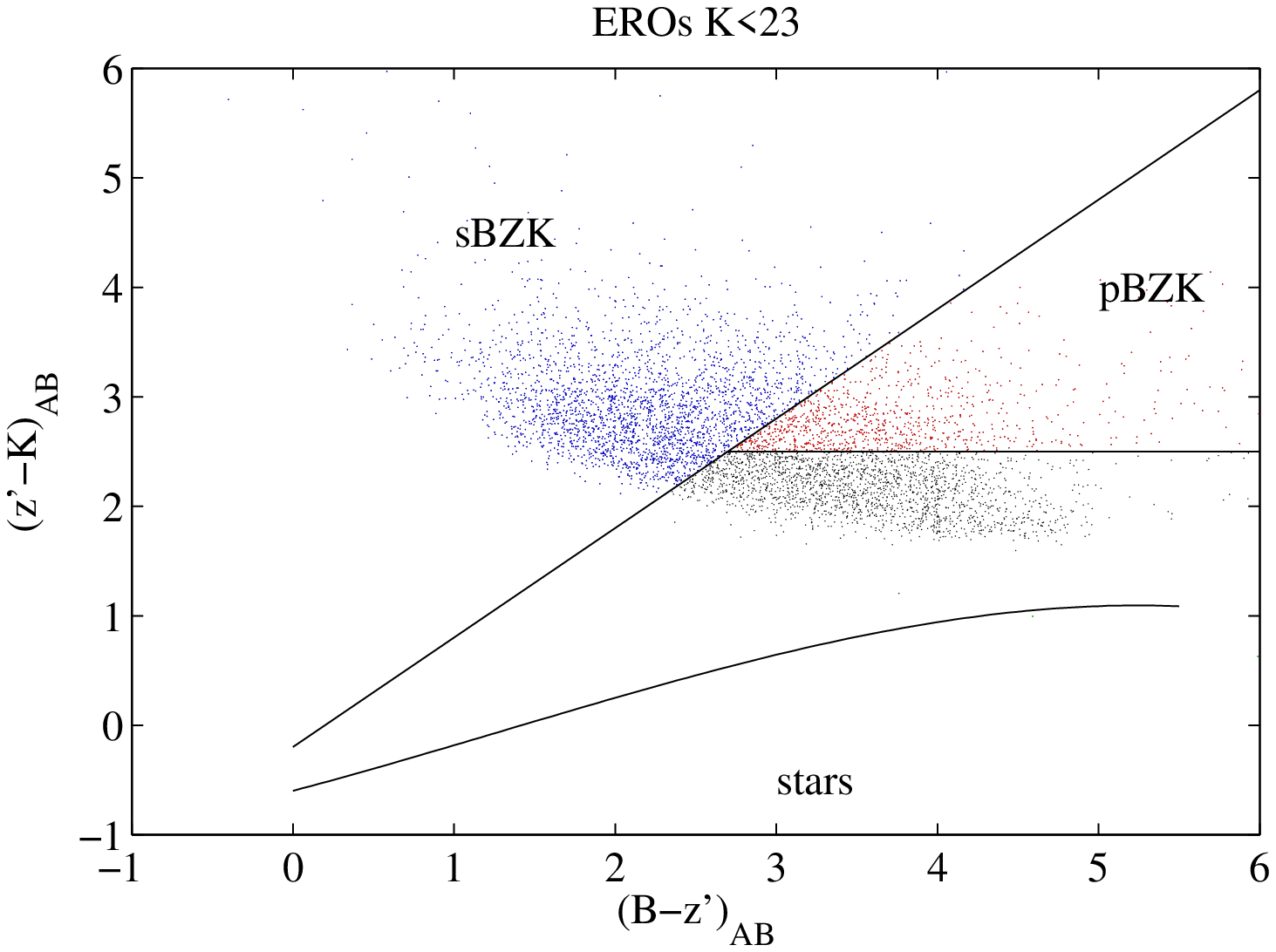}
\includegraphics[width=5.8cm]{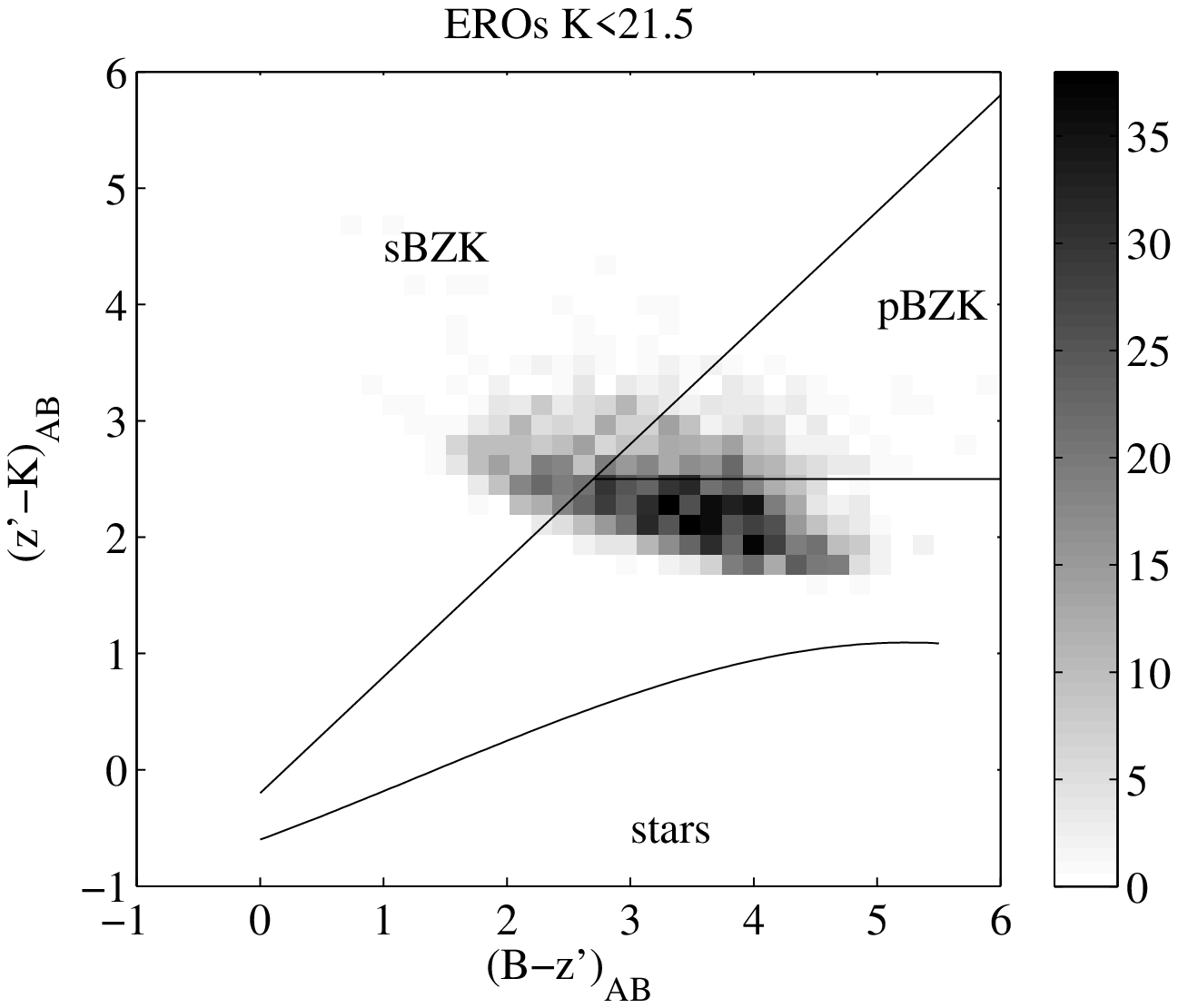}
\includegraphics[width=5.8cm]{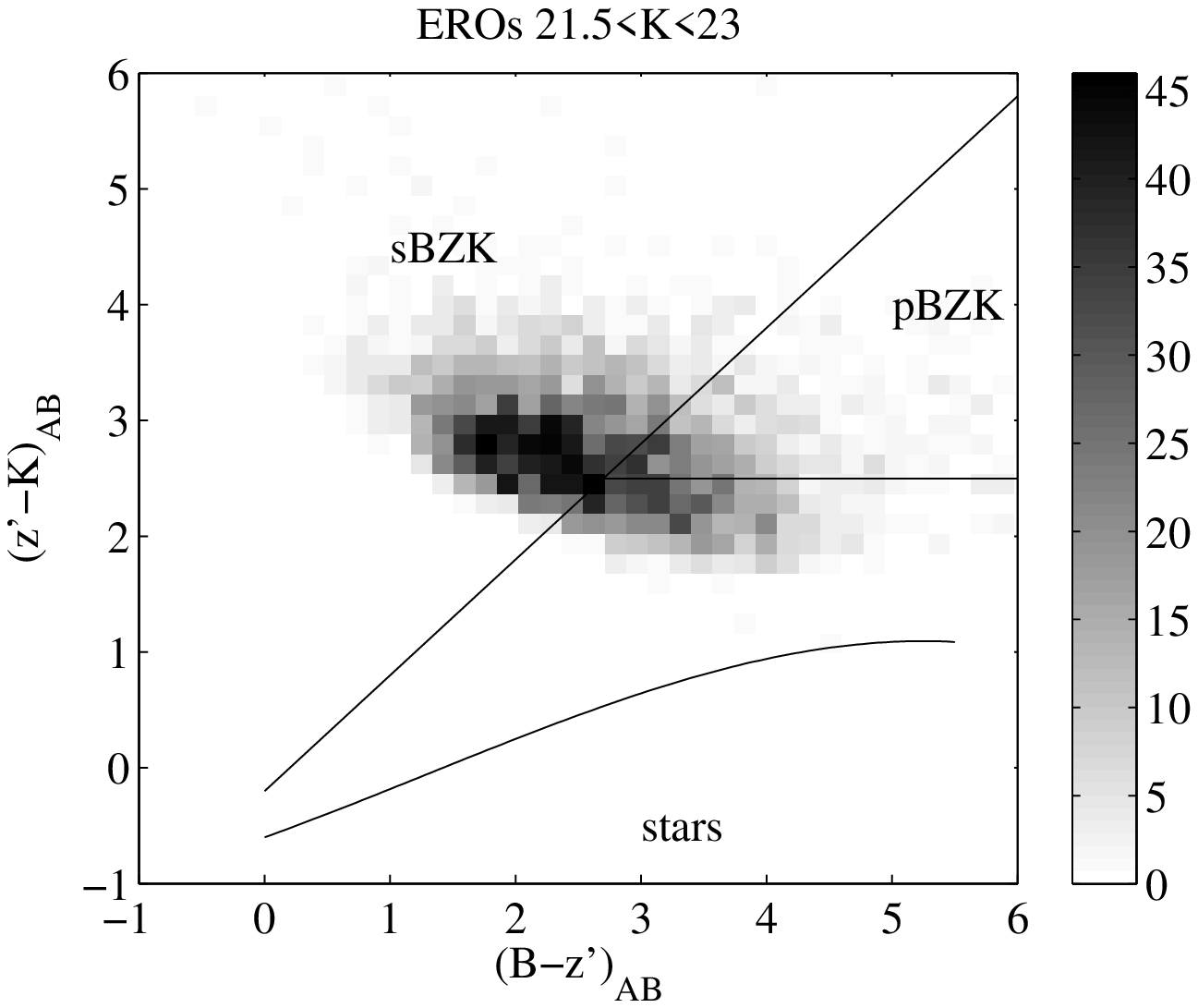}
\caption{\label{erobzk} Location of EROs on the BzK diagram.  The
position of the EROs {\em (left)} generally straddles the sBzK and
pBzK galaxies as well as the non-BzK region. Brighter EROs {\em
(middle)} are more likely to lie in the `passive' part of the diagram,
fainter EROS {\em (right)} are more likely to be sBzK galaxies.}
\end{figure*}


In Fig.~\ref{fluxKF} we show the correlation between $K$ mag and
\Smed\ for the full sample (top) and the BzK and ERO galaxies
(middle). There is a strong trend, with $S_{\rm 1,400} \propto
S_K^{0.53}$ for the full sample, and a steeper relationship for the
sBzK galaxies of $S_{\rm 1,400} \propto S_K^{0.79}$ and for the pBzK
galaxies of $S_{\rm 1,400} \propto S_K^{0.74}$. The difference in
slope is possibly due to the fact that, for the full sample, the
galaxies in the fainter $K$ bins have higher median redshifts than the
brighter bins. On the other hand, the BzK galaxies are selected by
colour to lie in a particular redshift range and so the median
redshifts are approximately the same for all $K$ bins. If the changes
in redshift are responsible for the different slopes then the flatter
slope indicates that the radio flux density is falling less quickly
with redshift compared to the $K$ flux. Fig.~\ref{fluxKF} shows that
the passive early-type pBzK galaxies have a lower radio/$K$-band flux
ratio than the star-forming sBzK galaxies. The fact that the slopes
are similar for the pBzK and sBzK samples suggests that the mechanism
responsible for the radio emission is similar for each class of galaxy
(or at least depends on $K$ in the same way). This is as one would
expect if pBzK samples contain a fraction of contaminating SFGs which
dominate their radio signal. We can follow this idea further by
plotting \Srad\ versus $K$ mag for subsets of pBzK galaxies, split by
redshift. pBzK galaxies at $z<1.4$ have larger 1,400-MHz fluxes and
luminosities than those at $1.4<z<2.5$, as shown in the bottom panel
of Fig.~\ref{fluxKF} where the black points are the sBzK sample, the
red points are the pBzK galaxies at $z<1.4$ and the blue points are
pBzK galaxies at $1.4<z<2.5$. It is now clear that the $z<1.4$ pBzK
follow the same \Srad\ versus $K$ mag relation as the sBzK galaxies,
supporting our hypothesis that they are star-forming interlopers in
pBzK colour space. The remaining `bona fide' pBzK galaxies at
$1.4<z<2.5$ are shown to have much less radio emission at a given $K$
mag than either sBzK or $z<1.4$ pBzK galaxies. There is not enough
signal to determine a slope to the high-redshift pBzK data. The
remaining radio flux from the high-redshift pBzK galaxies could come
either from a much lower-level contamination by SFGs, or from AGN. We
will investigate these two possibilities in \S\ref{AGN}.

The EROs are a heterogeneous sample. At brighter $K$ mags ($K > 21.5$)
they match the trends of the pBzK galaxies, but at fainter $K$ mags
the correlation between radio and $K$ seems to break down
(Fig.~\ref{fluxKF}-middle). In the BzK diagram (Fig.~\ref{erobzk}), the EROs
typically straddle the sBzK and pBzK colour range. Bright EROs have
redder $(B-z^{\prime})$ colours and so overlap more with the passive
region on the BzK plot. Fainter EROs are more aligned with the sBzK
part of the diagram. The point at which the EROs depart from the pBzK
relationship is similar to the magnitude of the break in the ERO
number counts at $K = 20.9-21.9$ (McCarthy et al.\ 2001; Smith et al.\
2002). This is further strong evidence that EROs are not a homogeneous
population, and the passive/star-forming split in an ERO sample will
depend strongly on the $K$-band selection limit for the sample. This
has been noted previously by Smail et al.\ (2002).


Various colour-selected samples have previously been stacked at
1,400\,MHz: BzK (Daddi et al.\ 2005) and EROs (Georgakakis et al.\
2006; Simpson et al.\ 2007; Ivison et al.\ 2007b). When we account for
the differing $K$ limits of the various samples using
Fig.~\ref{fluxKF} we find good agreement between our ERO and BzK
values and those in the literature. We will not discuss EROs further
as the BzK colour selection provides a cleaner separation of
high-redshift star-forming and passive galaxies.

\section{Reliability of radio emission as a tracer of star formation}
\label{AGN}

Integral to this paper is the assumption that we can use the 1,400-MHz
radio emission as a tracer of star formation. The main reason why this
may not be a valid assumption is the presence of radio-emitting AGN in
the sample at flux densities below the detection limit. Even after
removing the $z<1.4$ contaminants -- almost certainly SFGs -- we have
already seen that galaxies selected to be passive from the BzK diagram
have a residual level of radio emission which could be due to either
persistent, low-level contamination by SFGs or due to radio-emitting
AGN in pBzK galaxies. The sBzK galaxy population is known to consist
of star-forming galaxies and some AGN. Smol\v{c}i\'{c} et al.\ (2008)
show that Seyfert 2 galaxies at $z>1.4$ can lie in the sBzK region of
the diagram, while Daddi et al.\ (2004) and Reddy et al.\ (2005) have
both detected X-ray-bright AGN in the sBzK region. The contamination
rate from X-ray-emitting AGN was determined by Reddy et al.\ (2005) to
be 31 per cent for $19.9\leq K\leq 21.9$, 9 per cent for $21.9\leq
K\leq 22.4$ and 3 per cent for $22.4 \leq K\leq 22.9$. Reddy et al.\
also compared SFRs derived from X-ray stacking with and without the
detected X-ray sources and found their SFR estimates decreased by 13
per cent at $19.9\leq K\leq 21.9$, 3 per cent for $21.9\leq K\leq
22.4$ and 4 per cent for $22.4 \leq K\leq 22.9$. After excluding
X-ray-detected sources, their X-ray stack showed no hard-band
detection. They took this to indicate insignificant contamination from
low-luminosity AGN. Given that most of our sources (75 per cent) are
fainter in $K$ than the bins where Reddy et al.\ find significant AGN
contamination, we do not expect a problem if the radio emission trend
follows that of the X-rays.  The presence of optical-/X-ray-selected
AGN in our sample is not sufficient for us to assume that our radio
stacks are significantly contaminated by AGN-related emission. Seyfert
galaxies are known to follow the FIR/radio correlation for SFGs,
albeit with a greater scatter and a tendency to be slightly more radio
loud (Roy et al.\ 1998; Obri\'{c} et al.\ 2006; Mauch \& Sadler
2007). In fact, there are thought to be two radio classes of Seyfert
galaxy, one where the majority of the radio emission can be attributed
to star formation and another containing compact radio cores and
extended structures (`mini-lobes') which are radio loud and depart
more from the FIR/radio correlation than Seyferts without such radio
structures (Roy et al.\ 1998; Corbett et al.\ 2003). If our sBzK
sample is seriously contaminated by the radio-loud Seyferts then this
would show as a departure from the FIR/radio correlation (see
\S\ref{submm}). There has also been a comprehensive study of the radio
properties of optically-identified AGN from the SDSS survey by de
Vries et al.\ (2007), who stacked into the VLA FIRST survey. They
found that the radio emission in AGN samples which also show evidence
for star formation is dominated by star formation, with AGN
contributing at most 25 per cent of the total radio luminosity. They
find that the radio luminosity from `pure' AGN samples is a factor
4--5$\times$ lower at the same galaxy mass (velocity dispersion) than
that of AGN samples which also have ongoing star formation. If these
trends hold at higher redshift then we should not consider our sBzK
sample to be significantly contaminated by AGN emission. Magliocchetti
et al.\ (2008) studied the radio properties of bright, obscured
24-$\mu$m sources at $z=1-3$, classified into AGN and starburst
samples on the basis of their IRAC colours, finding a similar emission
mechanism in both samples, with the 1,400-MHz flux dominated by star
formation despite the presence of an AGN.

So, while the literature suggests we should not be unduly worried
about AGN contamination in this stacked sample, we are clearly in
uncharted waters as far as radio flux densities and redshift are
concerned.  We will now attempt to investigate the prevalence of AGN
in our stacked sample in two direct ways. First, we will look at the
radio spectral indices to see if they are commensurate with those
expected for a star-forming population. Second, we will compare the
radio-derived SFR to that measured using a different tracer; submm
flux density.

\subsection{\label{GMRT} Radio spectral indices}

\begin{table*}
\caption{\label{GMRTflux} The 610-MHz flux densities for $K$-selected
populations and derived spectral indices.}
\begin{tabular}{lccccc}
\hline
\multicolumn{1}{c}{Sample}&\multicolumn{1}{c}{N}&\multicolumn{1}{c}{\%
det}&\multicolumn{1}{c}{\Sgmrt}&\multicolumn{1}{c}{\Smedgmrt}&\multicolumn{1}{c}{$\alpha^{610}_{1,400}$}\\
\multicolumn{3}{c}{}&\multicolumn{1}{c}{($\mu$Jy)}&\multicolumn{1}{c}{($\mu$Jy)}&\multicolumn{1}{c}{}\\
\hline
All ($z>0.2$) & 36 057 & 0.6 & $13.8\pm 0.3$ & $13.0\pm 0.5$ & $-0.74\pm0.03$\\
Non-BzK & 24 772 & 0.6 & $12.6\pm 0.4$ & $12.0\pm 0.5$ & $-0.78\pm 0.05$\\
sBzK & 10 534 & 0.8 & $16.8\pm 0.6$ & $15.8\pm 0.8$ & $-0.72\pm 0.05$ \\
pBzK & 857 & 0.8 & $12.3\pm 2.1$ & $12.8\pm 3.0$ & $-0.60\pm 0.24$ \\
ERO & 4552 & 1.4 & $19.3\pm 0.9$ & $18.5\pm 1.5$ & $-0.82\pm 0.07$
\\
\hline
\end{tabular}
\flushleft
\footnotesize{Notes: (1) Sample name; (2) number in stack (including
detections); (3) fraction detected at $\geq$5\,$\sigma$; (4)
noise-weighted mean excluding detections; (5) median and 68-per-cent
CI; (6) spectral index from 610 to 1,400-MHz, where $S_{\nu} \propto
\nu^{\alpha}$. All spectral index values are quoted using the
integrated mean values,
there is no significant difference when using the median. Errors on
$\alpha$ are propagated from the error in the mean stacked flux
densities at 610 and 1,400-MHz}
\end{table*}

\begin{figure*}
\includegraphics[width=4.3cm]{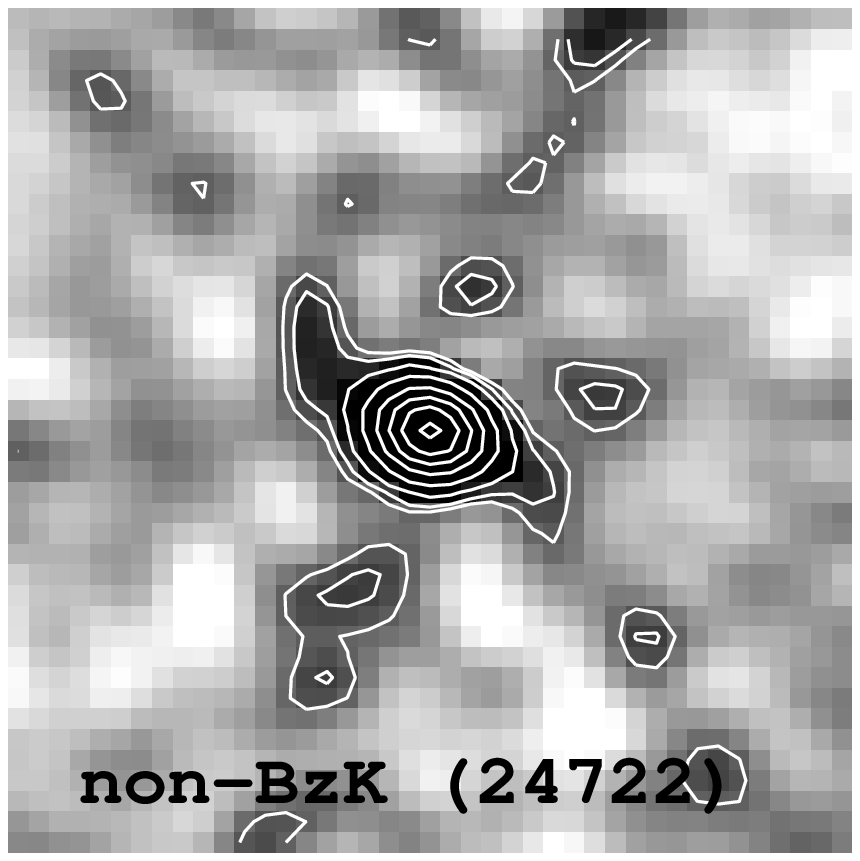}
\includegraphics[width=4.3cm]{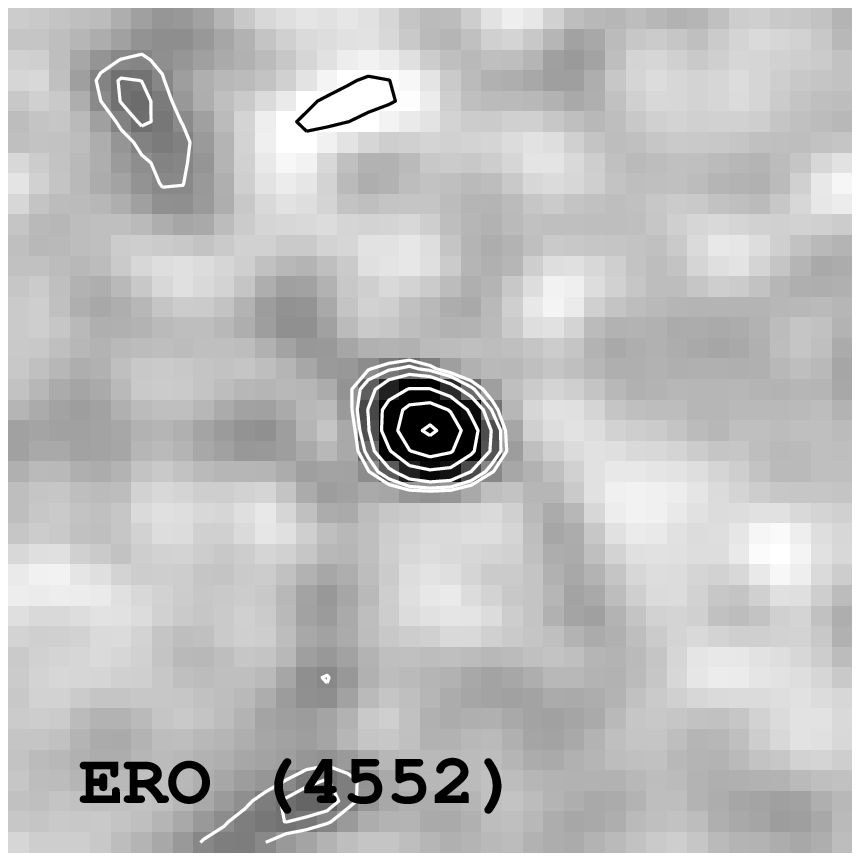}
\includegraphics[width=4.3cm]{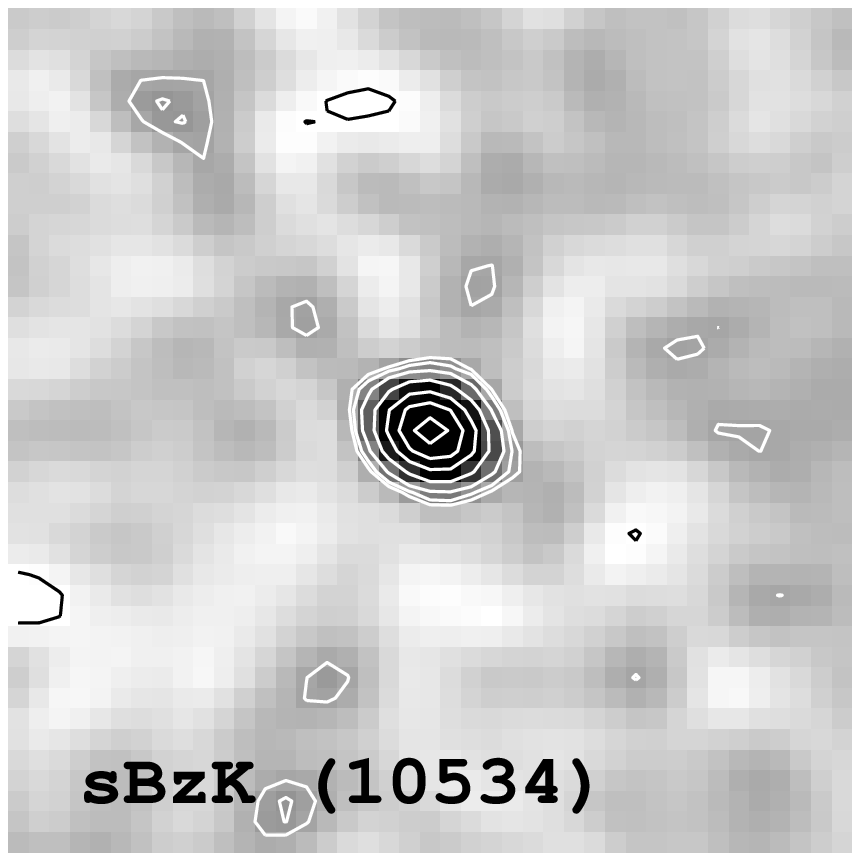}
\includegraphics[width=4.3cm]{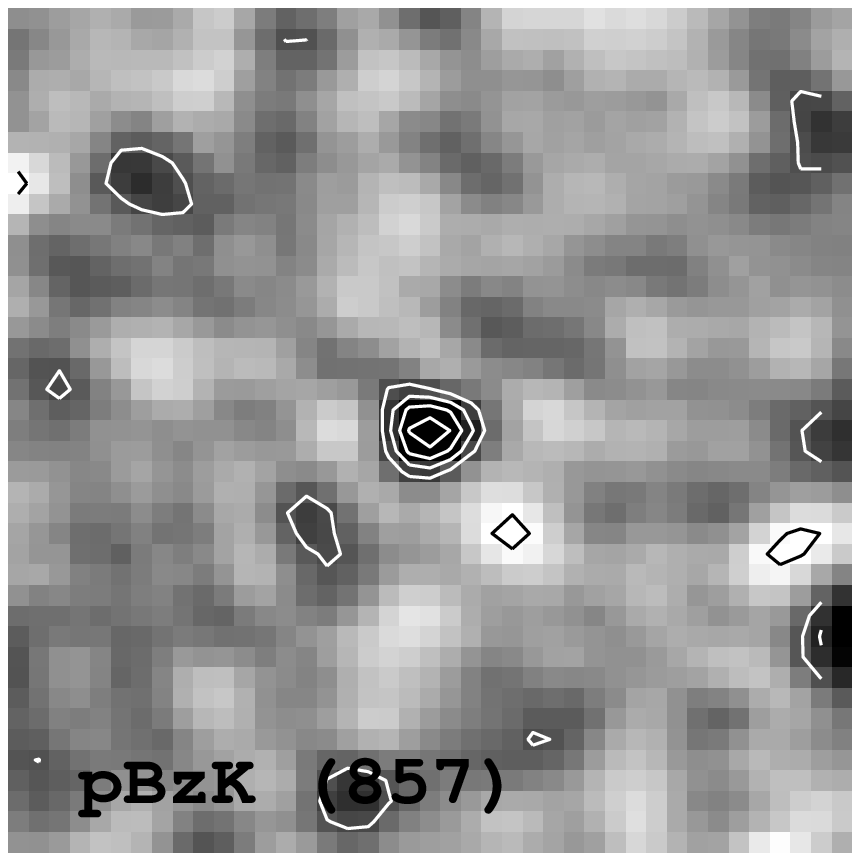}
\caption{\label{GMRTpostage} Postage-stamp images of the GMRT 610-MHz
stacks for non-BzK, ERO, sBzK and pBzK samples. Contours for the
non-BzK, ERO and sBzK are $-3,3,4,6,8$ then in steps of $+2\sigma$
while for pBzK they are $-2, 2,3,4$ then in steps of $+1\sigma$. Stacks are the
weighted-mean with $\pm 5\sigma$ pixels removed before stacking.}
\end{figure*}

The spectral indices of the mJy and sub-mJy radio populations can give
some clues to the nature of the radio emission. Above 1\,mJy,
steep-spectrum, radio-loud AGN dominate with $\alpha \sim -0.7$ (Bondi
et al.\ 2007). Below 1\,mJy, where the radio counts flatten and the
new population emerges, the spectral index has been found to flatten
(Ciliegi et al.\ 2003; Prandoni et al.\ 2006). Bondi et al.\ (2007)
conducted a deep survey of the VVDS-VLA field with the GMRT and found
that while the spectral index flattens from $\alpha=-0.67\pm 0.05$ at
$S_{\rm 1,400}>500\, \mu$Jy to $-0.46\pm 0.03$ from $150 < S_{\rm
1,400} < 500\, \mu$Jy, it then steepens to $\alpha=-0.61\pm 0.04$ at
fainter fluxes. They also find that early-type galaxies have flatter
indices ($\alpha=-0.55\pm0.04$) than late-type/starburst galaxies
($\alpha\sim -0.7$). Fomalont et al.\ (2006) compared 1,400- and
8,400-MHz maps and found that sources with $S_{\rm 1,400}<75\, \mu$Jy
have steeper indices than brighter sources ($\alpha=-0.87\pm 0.05$,
compared to $\alpha=-0.78\pm 0.04$). They claim that the AGN fraction
is therefore decreasing below $100\, \mu$Jy. While the flattening of
the spectral index between $200 <S_{\rm 1,400} < 1,000\,\mu{\rm Jy}$
was thought to be due to an increase in the contribution from
flat-spectrum compact radio cores hosted by low-luminosity AGNs, with
the subsequent steepening at $S_{\rm 1,400} < 200\ \mu$Jy ascribed to
an increasing dominance of starburst/late-type
galaxies. Unfortunately, this observed steepening at sub-mJy fluxes is
not yet a statistically robust phenomenon, due partly to the {\em
spectral index bias:} a population selected at a high frequency is
less likely to be found to have flat spectral indices if the
low-frequency flux limit is higher than the high-frequency limit.

We will now investigate whether stacking the $K$-selected populations into
radio maps at two frequencies can shed any light on the behaviour of the
radio spectral index at $S_{\rm 1,400}<50\ \mu$Jy.

\subsubsection{GMRT observations and stacking}

For our second frequency, we employ a 610-MHz map taken with the Giant
Metre-wave Radio Telescope (GMRT\footnote{We thank the staff of the
GMRT that made these observations possible. GMRT is run by the
National Centre for Radio Astrophysics of the Tata Institute of
Fundamental Research.}) near Pune, India. Data were obtained in the
UDS field during 2006 February 03--06 and December 05--10, primarily
to explore the spectral index of the radio emission from
submm-selected galaxies, though a byproduct has been a thorough
exploration of spectral indices of the faint background population
(Ibar et al., in preparation). Typically, we were able to employ
27--28 of the 30 antennas that comprise the GMRT, observing a mosaic
of three positions, arranged like the VLA mosaic. The total
integration time in each field, after overheads, was around 12\,hr. We
recorded 128$\times$1.25-kHz channels every 16\,s in the lower and
upper sidebands (602 and 618\,MHz, respectively), in each of two
polarisations.  Integrations of 40-min duration were interspersed with
5-min scans of the bright nearby calibrator, 0240$-$231, with scans of
3C\,48 and 3C\,147 for flux and bandpass calibration.

Calibration initially followed standard recipes within \AIPS. However,
because of concerns that some baselines were picking up signal from
local power lines, a raft of new measures were introduced to avoid
detrimental effects on the resultant images. These are to be reported in
detail in Ibar et al.

Imaging each of these datasets entailed mosaicing 37 facets, each
512$^2\times$ 1.25$^2$-arcsec$^2$ pixels, to cover the primary beam. A
further 6--12 bright sources outside these central regions, identified
in heavily tapered maps, were also imaged. Our aim was to obtain the
best possible model of the sky. {\sc clean} boxes were placed tightly
around all radio sources for use in self-calibration, first in phase
alone, then in amplitude and phase.

The final six mosaics, two for each pointing (upper and lower
sidebands), were convolved to a common beam size, then knitted
together using {\sc flatn}. An appropriate correction was made for the
shape of the primary beam and data beyond the quarter-power point of
the primary beam were rejected.  Data for each pointing and sideband
were weighted according to individual noise levels. The final {\sc
flatn}ed image has a noise level of 57\,$\mu$Jy\,beam$^{-1}$ ---
significantly higher than images of the Lockman Hole
field during the same observing period. The reasons for this include
the extra RFI suffered during day-time observing, the low declination
of the UDS and the presence of several very bright radio emitters in
the field.

The stacking was performed in the same way as for the 1,400-MHz map,
except that the BWS correction is negligible for the GMRT's narrow
channels. Regions with $\sigma_{\rm 610MHz} \leq 120\,\mu$Jy were
included in the stack. We calculated the correction for integrated
flux and dirty-to-{\sc clean} as we have for the 1,400-MHz data. The
final correction value was found to be a factor $1.20\pm 0.15$. All
spectral indices are quoted after making corrections at both
frequencies.

Due to the noise being much higher in the GMRT data, only stacks of
total populations could be made (not split as a function of redshift,
$K$ mag, etc.). We also stacked sets of random positions into the GMRT
map to check for systematics and we found that the means are
consistent with zero. The medians are generally a little below the
means and there is a possible $-0.1\ \mu$Jy offset, consistent with
the median background removal applied to the map which would have
resulted in negative patches around bright sources. We will not
correct for this effect as it is negligible.

\subsubsection{Spectral index results}

The results of the stacking for the $K$-selected populations are
listed in Table~\ref{GMRTflux} and the postage-stamp images are shown
in Fig.~\ref{GMRTpostage}. It must be noted here that our spectral
index measurements are the ratio of two `average' quantities, flux at
610\,MHz and flux at 1,400\,MHz. As such, we cannot calculate the true
average spectral index, as would be the case if samples of sources
were detected at both wavelengths and their individual spectral index
values averaged. It is possible that there are situations where this
may bias our values of the spectral index, particularly if $\alpha$ is
correlated with \Srad\ as seems to be suggested in the literature
(Fomalont et al.\ 2006; c.f. Ibar et al.\ in preparation). We
investigated this possibility using a simple Monte-Carlo
simulation. We generated sources with a power-law distribution of
$S_{1,400}$, following $dN = S^{-\gamma}\,dS$ and for each source we
assigned a value of $\alpha$ between $-0.9$ and $0.0$. We provided the
$\alpha$ values according to two models, one where $\alpha$ is random
and another where $\alpha$ depends on $S_{1,400}$ as $\alpha=-0.9+0.5
\log\,(1+ $\Srad$/50)$. Below 50\,$\mu$Jy which we
introduce an asymptote to $\alpha=-0.9$ as \Srad\ approaches zero. These
choices reproduce the mean fluxes and values for $\alpha$ we find for
our stacks. We then used the values of $S_{1,400}$ and $\alpha$ for
each artificial source to calculate the value of $S_{610}$. The
average spectral index in the simulation was then compared to the
spectral index derived from the mean fluxes at both frequencies, as we
have done in the stacks. The results showed that for input source
counts with slopes $\gamma > 1.5$, the bias on the spectral index
derived from the stacking method is very small -- a flattening of 0.01
in $\alpha$ for the model where $\alpha$ varies with $S_{1,400}$ and a
steepening in $\alpha$ of 0.05 for the model where $\alpha$ is
random. The slope of the 1,400-MHz differential counts at faint fluxes
$\sim$30--75\,$\mu$Jy is 2.0--2.5 (Fomalont et al.\ 2006; Biggs \&
Ivison 2006). If this holds at fainter fluxes, our method should not
significantly bias the measured spectral indices; indeed, it
probably gives values closer to the true spectral index than surveys
which use detected sources.

The spectral indices are consistent, within the errors, across all the
samples. 
The very faint flux regimes we are probing here are below any other
work which has measured spectral indices.
The average for the whole $K$-selected sample at
($z>0.2$) is $\alpha=-0.74\pm0.03$. 
This is consistent with values for $\alpha$ in star-forming samples
(Bondi et al. 2007; Ibar et al. in preparation) but is lower than the
value found at faint fluxes by Fomalont et al. 2006. It may have been
expected that pBZK galaxies would show a latter value for $\alpha$
then sBZK galaxies if radio-quiet AGN were the dominant contributors
to the stacked radio flux. There is a hint that such a flattening is
present but, with the current noise levels on the pBZK stack, this
difference is not significant.

 The corrections we have had
to make for integrated flux versus peak flux mean that there are
further systematic errors ($\sim \pm0.15$) 
atop those quoted (which are from the stack
statistics only). The GMRT beam is particularly unpleasant in the UDS
field due to poor $(u,v)$ coverage, meaning the correction factor to
{\sc clean}-integrated flux density is more uncertain than that
derived at 1,400\,MHz, where it was clear that source size and
positional errors were producing a broadening compared to the VLA
beam. Although we can compare samples in our data in a consistent way
and are confident that they are as similar or different as they
appear, the {\em exact value\/} of the spectral index is rather
sensitive to the corrections we are making. On the other hand, we are
not hampered by the {\em spectral index bias}, as mentioned above,
since we do not need to detect sources in both catalogues in order to
measure the statistically averaged indices. Overall we would argue
that this spectral index investigation is 
consistent with the hypothesis that star formation is
the main contributor to the radio emission in these stacks as implied
by Fomalont et al.\ (2006), Seymour et al.\ (2008) and Ibar et al.\
(in preparation).

\subsection{Submm stacking of BzK galaxies}
\label{submm}

As another test of the reliability of the radio as a tracer of star
formation, we compared it to the SFR from stacking the sBzK and pBzK
samples into the submm map of the SXDF from the SHADES survey (Mortier
et al.\ 2005; Coppin et al.\ 2006). Noise-weighted means and medians
were calculated at the pixels corresponding to the positions of the
$K$ galaxies. For the mean stacks, we excluded pixels with
$\rm|S/N|\geq 5$. The submm beam is 14\,arcsec and so the stacked
pixel fluxes should represent the total flux for the $K$ galaxies. To
check that there were no systematic effects, we examined the
distribution of pixel signal-to-noise ratios (SNRs)
 in the map, as seen in figure 9 of Mortier
et al.\ (2005). The mean SNR of 0.04 and variance of 1.15, which for a
mean noise level in the map of $2.2\, \rm{mJy\,beam^{-1}}$, are
consistent with zero.

\begin{figure}
\hspace{1.5cm}
\includegraphics[width=7cm]{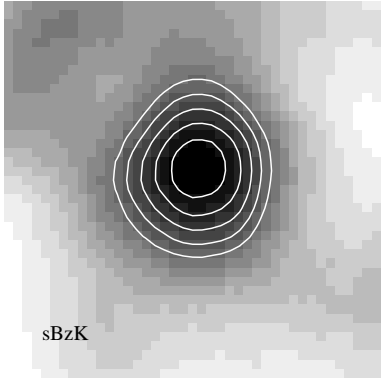}
\caption{\label{submm_sn} Stacked image of the 850$\mu$m SHADES map at
the locations of the sBzK sources (1421 in total, excluding the one
detection at $>5\sigma$). Contours are at $5,6,7,8,9 \,\sigma$ and the
image is $41\times 41$-arcsec.}
\end{figure}

The resulting 850-$\mu$m stacked flux was extrapolated to $L_{\rm
FIR}$ using a modified black-body SED ($\nu^{\beta}\, B(\nu,T)$) with
a dust emissivity index $\beta=1.5$ and dust temperatures of 30\,{\sc
k} and 35\,{\sc k}. Two reasonable temperatures were chosen to provide
an indication of the effect temperature has on the estimate of $L_{\rm
FIR}$. $L_{\rm FIR}$ was then converted to an SFR using the
relationship given by Kennicutt (1998), where ${\rm SFR} (M_{\odot}\,
\rm{yr^{-1}}) = 1.7\times 10^{-10} \, L_{\rm FIR}/L_{\odot}$. Only one
pixel in the sBzK galaxy stack was at $>$5$\sigma$ and so we are
confident that the noise-weighted mean is the most efficient estimator
of the average properties of the submm stack. The results are
presented in Table~\ref{submmT} and show that the radio- and
submm-derived SFRs are in good agreement, within the uncertainties of
the calibrations. This is another reassurance that the radio emission
from the sBzK galaxy sample is not dominated by AGN. The overall
detection of sBzK galaxies is in good agreement with that of Takagi et
al.\ (2007) who found a mean $S_{\rm 850\mu m}=0.53\pm 0.19$\,mJy,
though with far fewer sources (112 compared to 1,421). It is
interesting to note that the 850-$\mu$m stacked flux density also
correlates with absolute $K$ mag.

The total pBzK sample was not detected and the small number of sources
overlapping the submm map means that the upper limit is not sensitive
enough to constrain the nature of the radio emission. The upper limit
to the submm-derived SFR is consistent with that determined through
radio stacking. On splitting the pBzK galaxies into two redshift bins,
$z<1.4$ and $z>1.4$, we see a 2.6-$\sigma$ detection in the submm band
in the low-redshift bin. Although not formally significant, this
suggests that the $z<1.4$ pBzK galaxies are a strongly star-forming
sample, and their submm emission is consistent with radio emission
powered by star formation. At $z>1.4$, there is no submm emission, but
again the upper limits are not deep enough to constrain the origin of
the radio emission.

Another way to look at the stacked submm and radio fluxes is in terms
of the FIR/radio correlation. The FIR/radio ratio is defined as 
\[
q = \log[\frac{S_{\rm FIR}/3.75\times 10^{12} \,\rm Hz}{S_{\rm
rad}}]
\]
where $S_{\rm FIR}$ is given by Helou et al.\ (1985) as
\[
S_{\rm FIR} = 1.4\times 10^{-14}(2.58\, S_{60} + S_{100}) \, {\rm W
m^{-2}}
\] 
where $S_{60}$ and $S_{100}$ are the 60- and 100-$\mu$m flux densities in
Jy. $S_{\rm rad }$ is the radio flux in units of ${\rm W m^{-2}
Hz^{-1}}$ and several different frequencies are used in the literature. 

Radio-bright AGN (such as core-dominated Seyferts) have lower $q$
values than samples composed purely of SFGs. We have calculated $q$
for our sample of sBzK galaxies under the two temperature assumptions
used above (since we do not measure directly the rest-frame 60- and
100-$\mu$m part of the spectrum). The results are summarised in
Table~\ref{radio/FIR}. Unfortunately, our lack of knowledge of the
dust temperatures in the sBzK galaxies limits the usefulness of this
approach. For a dust SED with $T=30$\,{\sc k}, we find $q$ is low
compared to the star-forming FIR/radio relation, and more similar to
the values found in Seyfert samples; for a dust SED with $T=35$\,{\sc
k}, we find $q$ to be entirely consistent with star-forming
values. The average temperature of {\em IRAS}-bright galaxies in the
local Universe is $T\sim 35$\,{\sc k} (Dunne et al.\ 2000), and
galaxies containing an AGN would usually tend to have hotter dust than
those without -- due to extra heating of dust near the nuclear
regions. The $q$ values we find are only similar to those of Seyferts
for the coldest dust temperature assumptions, if we use a dust
temperature more appropriate for AGN dominated galaxies
(e.g. $T=50$\,{\sc k}; Kuraszkiewicz et al.\ 2003) the we see from
Table~\ref{radio/FIR} that the $q$ values are then inconsistent with
those derived for AGN. This is further evidence that star formation is
dominating the radio emission of our sBzK sample. Further
investigation of the FIR/radio relationship will be possible with
forthcoming {\em Herschel} observations which will directly measure
the temperature of the dust out to high redshift.

\begin{table*}
\caption{\label{submmT} Results of stacking BzK galaxies into the
SHADES 850-$\mu$m SXDF map.}
\begin{tabular}{lcccccccc}
\hline
\multicolumn{1}{c}{Sample}&\multicolumn{1}{c}{$K_{\rm
abs}$(Vega)}&\multicolumn{1}{c}{$\langle z\rangle$}&\multicolumn{1}{c}{N}&\multicolumn{1}{c}{$\bar{S}_{\rm 850\mu m}$}&\multicolumn{1}{c}{$\rm
SFR^{sub}_{35K}$}&\multicolumn{1}{c}{$\rm
SFR^{sub}_{30K}$}&\multicolumn{1}{c}{$\rm
SFR^{rad}_{C}$}&\multicolumn{1}{c}{$\rm SFR^{rad}_{B}$}\\
\multicolumn{4}{c}{}&\multicolumn{1}{c}{(mJy)}&\multicolumn{1}{c}{($\rm
M_{\odot}\,yr^{-1}$)}&\multicolumn{1}{c}{($\rm
M_{\odot}\,yr^{-1}$)}&\multicolumn{1}{c}{($\rm
M_{\odot}\,yr^{-1}$)}&\multicolumn{1}{c}{($\rm
M_{\odot}\,yr^{-1}$)}\\
\hline
sBzK & all & 1.56 & 1,421 & $0.53\pm 0.06$ & $156.3\pm 16.6$ & $80.2\pm
8.9$ & $153.6\pm 6.7$ & $70.6\pm 3.1$\\  
     & $-(21.5 - 23.5)$ & 1.41 & 187 & $0.17\pm0.16$ & $<128$ & $<65$ & $30\pm 12$ & $14\pm 6$\\
     & $-(23.5 - 24.5)$ & 1.54 & 509 & $0.17\pm 0.10$ & $<83$ & $<42$ &
$92\pm 10$ & $42\pm 5$ \\
     & $-(24.5-25.5)$ & 1.63 & 442 & $0.76\pm0.10$ & $230\pm 31$ &
$119\pm 16$ & $226\pm 11$ & $104\pm 5$ \\
     & $-(25.5-26.6)$ & 1.74 & 212  & $1.19\pm 0.15$ & $380\pm 46$ &
$197\pm 24$ & $485\pm 27$ & $223\pm 12$\\
     & $-(26.5-27.5)$ & 2.16 & 33 & $1.60\pm 0.37$ & $618\pm 141$ &
$329\pm 75$ & $1267\pm 175$ & $ 583\pm 80$ \\
\hline
pBzK & all & 1.53 & 147 & $0.22\pm 0.18$ & $<158$  & $<81$  & $131\pm 19$  & $61\pm 9$ \\
pBzK $z<1.4$ & all & 1.28 & 46 & $0.89\pm 0.34$ & $219\pm 84$ &
$111\pm 42$  & $158\pm 19$ & $73\pm 9$ \\
pBzK $z>1.4$ & all & 1.66 & 101 & $-0.05\pm 0.22$ & $<200$ & $<103$ &
$105\pm 31$ & $48\pm 14$\\
\hline
\end{tabular}
\flushleft
\footnotesize{Columns: (1) Sample name; (2) absolute $K$ mag; (3)
redshift; (4) number in submm stack; (5) mean noise-weighted flux at 850\,$\mu$m; (6) SFR
derived from submm flux density, using a dust SED with $\beta=+1.5$ and $
T=35$\,{\sc k}; (7) as for (6) but using $T=30$\,{\sc k}; (8) SFR derived from
the 1,400-MHz flux density using the Condon SFR conversion; (9) as for (8) but
using the Bell SFR conversion.} 
\end{table*}

\begin{table*}
\caption{\label{radio/FIR} Comparison of FIR/radio ratios for sBzK and
samples from the literature.}
\begin{tabular}{cccccc}
\hline
\multicolumn{1}{c}{Ref.}&\multicolumn{1}{c}{$q_{\rm
SF}$}&\multicolumn{1}{c}{$q_{\rm AGN}$}&\multicolumn{1}{c}{$q_{\rm
sBzK}$}&\multicolumn{1}{c}{$q_{\rm sBzK}$}&\multicolumn{1}{c}{$q_{\rm sBzK}$}\\
\multicolumn{1}{c}{}&\multicolumn{1}{c}{}&\multicolumn{1}{c}{}&\multicolumn{1}{c}{(30K)}&\multicolumn{1}{c}{(35K)}&\multicolumn{1}{c}{(50K)}\\ 
\hline
A & $2.83\pm 0.03$ & $2.55\pm0.14$ & $2.53\pm0.05$ & $2.82\pm0.05$ & $3.36\pm0.05$\\
B & $2.48\pm0.10$ & $1.87\pm 0.08$ & $2.18\pm0.05$
& $2.48\pm0.05$ & $3.02\pm0.05$\\
C & $2.300\pm 0.003$ & $2.00\pm 0.04$ & $2.00\pm0.05$ & $2.29\pm0.05$
& $2.82\pm 0.05$ \\
\hline
\end{tabular}
\flushleft
\footnotesize{Columns: (1) Literature samples. A: Corbett et al.\ (2003) at 4.8\,GHz, B: Roy et
al.\ (1998) at 2.4\,GHz, C: Mauch \& Sadler (2007) at 1.4\,GHz; (2)
$q$ value found for SFGs in the literature samples; (3) $q$ value
found for AGN in the literature samples; (4) $q$ value from sBzK submm
and radio stacks assuming a dust temperature of 30\,{\sc k}, scaled to
the appropriate frequency using $\alpha=-0.8$. (5) $q$ value from sBzK
submm and radio stacks assuming a dust temperature of 35\,{\sc k},
scaled to the appropriate frequency using $\alpha=-0.8$. (6) $q$ value
from sBzK submm and radio stacks assuming a dust temperature of
50\,{\sc k}, scaled to the appropriate frequency using $\alpha=-0.8$.}
\end{table*}
 
\section{Discussion: the SFRs of $K$-selected galaxies}

We now split the samples into bins of redshift and stellar mass
($M_{\rm stellar}$), or absolute rest-frame $K$ mag. The rest-frame
$K$ mags are calculated with the photometric redshifts using the
best-fitting SED template. These are then converted into $M_{\rm
stellar}$ using the relationship given by the Millennium simulation
(de Lucia et al.\ 2006) in the manner employed by Serjeant et al.\
(2008) which is effectively using a Salpeter IMF. This is a
redshift-dependent conversion which agrees with observational data
from the Munich Near-IR Cluster Survey (MUNICS -- Drory et al.\
2004). There are, of course, uncertainties in moving between $K$ and
$M_{\rm stellar}$; the scatter in the relationship used here is of
order 0.2-0.3 dex at a given redshift. A more careful measurement will
be possible in the future using the results of the {\em Spitzer}
Legacy survey of the UDS, in particular the 3.6--8-$\mu$m data. We
have checked that using $M_{\rm stellar}$ rather than the simpler
absolute $K$ mag does not change the trends discussed in this
section. We use $M_{\rm stellar}$ as this is more directly comparable
to other work in the literature.

\subsection{Star-formation history}

\begin{figure}
\includegraphics[width=8.8cm]{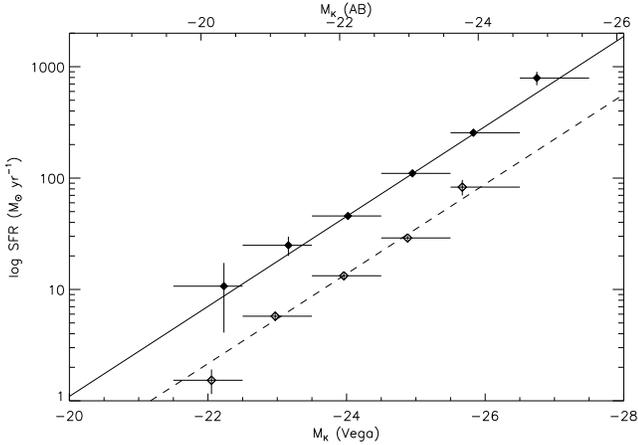}
\caption{\label{sfrKabsF} Radio-derived SFR versus absolute $K$ mag
for $z<1$ (open diamonds) and $z>1$ (filled diamonds). The best fits
are shown and are consistent with ${\rm SFR} \propto L_{\rm K}$.}
\end{figure}

\begin{figure*}
\includegraphics[width=8.8cm]{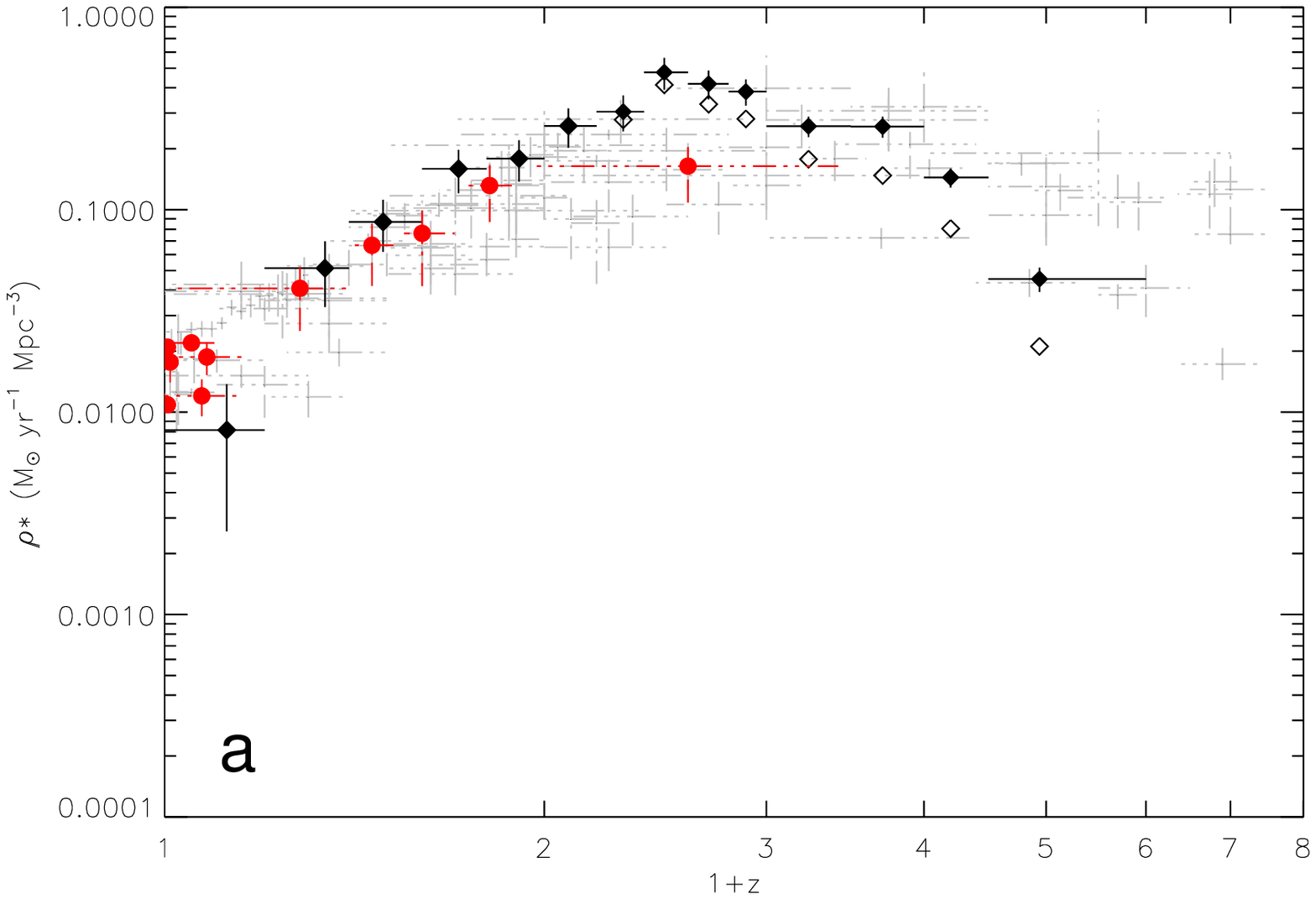}
\includegraphics[width=8.8cm]{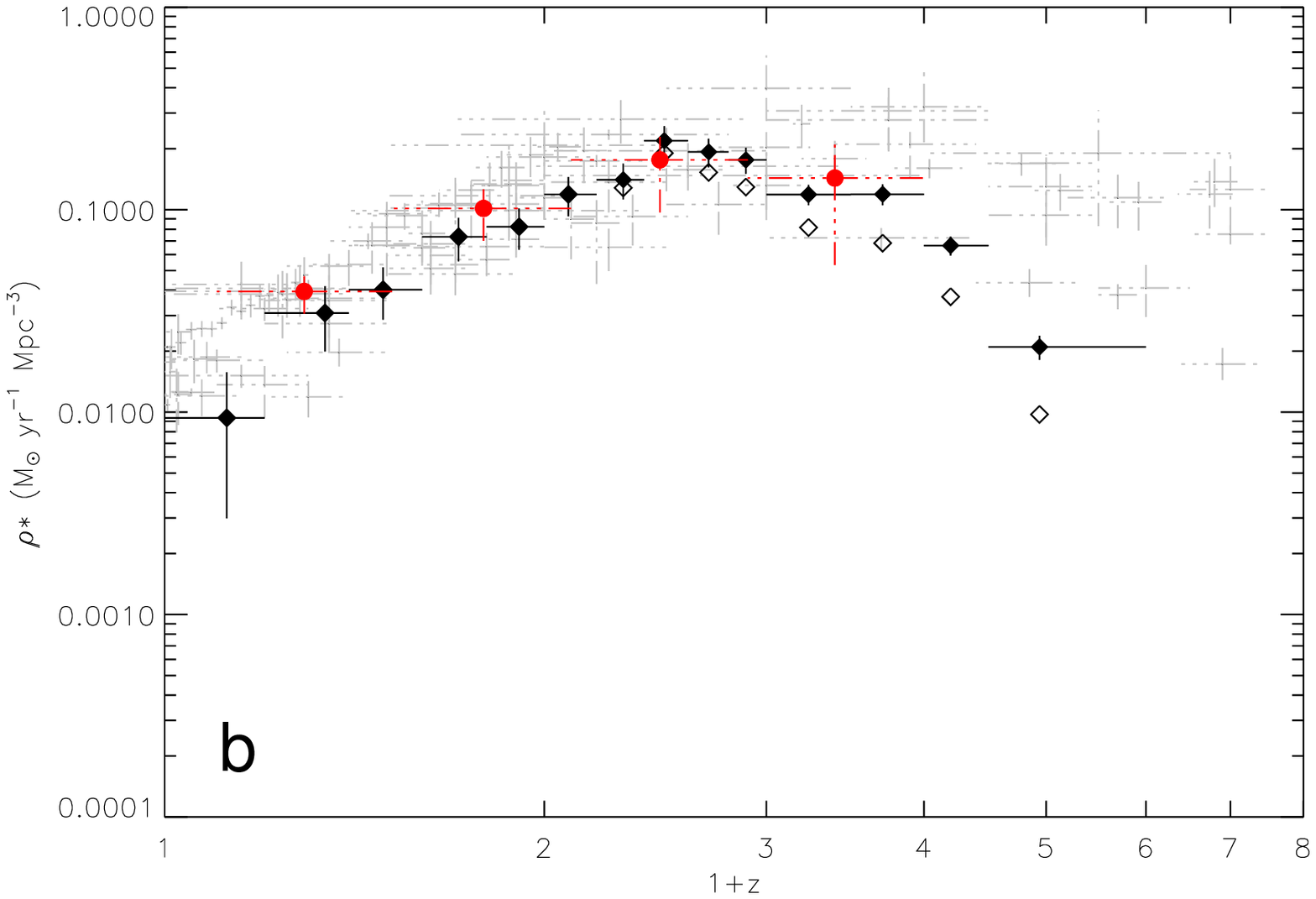}
\caption{\label{SFRD} Radio-derived star-formation histories using:
(a) the SFR relationship of Condon (1992); (b) that of Bell
(2003). The faint points are the compilation of HB06 normalised to a
Salpeter IMF. Red circles in (a) show literature values (Condon 1989;
Condon et al.\ 2002; Haarsma et al.\ 2000; Machalski \& Godlowski
2000; Sadler et al.\ 2002; Mauch \& Sadler 2007) derived from
1,400-MHz data using the Condon (1992) conversion. Red circles in (b)
show the values from Seymour et al.\ (2008) at 1,400\,MHz for the Bell
(2003) SFR conversion, converted to a Salpeter IMF. All the SFRD points are
corrected for the detection incompleteness, filled symbols are also
corrected for LF incompleteness using the UDS LF from Cirasuolo et
al.\ (2008) while open symbols show LF uncorrected points. Out to $z\sim
2$, the radio-derived points using both SFR relations are in good
agreement with literature values, both in the radio and other optical
wavelengths. The radio-derived SFRD then declines more steeply than
the points from HB06.}
\end{figure*}

To begin with, we look at the SFRD as a function of redshift for our
full sample (i.e. `All galaxies' from Section~\ref{IRsel}). The sample
was divided into a number of redshift bins and the median SFR for each
bin calculated. The volume contained within each bin was calculated
using the effective survey area (UDS/SuprimeCam/VLA overlap field,
excluding regions around bright stars and with $\sigma_{\rm 1,400}
\leq 20\,\mu$Jy) which is 1311.1\,arcmin$^2$. The space density is the
number of sources meeting the selection criteria in that redshift bin
divided by the volume in the bin. The space density is corrected for
the 80 percent completeness of the $K$-sample in the magnitude range
$22.5<K\leq 23.0$ by finding the number of galaxies within each
redshift bin in this magnitude range, multiplying that number by 1.2
and adding it to the number of galaxies in that redshift bin at $K\leq
22.5$. These corrections are small and range from 1.04 at low redshift
to 1.13 at $z>3.0$. The SFRD is estimated by multiplying the space
density by the median stacked SFR for each redshift bin. The results
are shown in Fig.~\ref{SFRD} for both methods of estimating the
SFR. The errors on the points are the sum in quadrature of the error
on the median SFR for a bin, the Poisson $1/\sqrt{N}$ contribution and
cosmic variance errors, derived from fig.~2 in Somerville et al.\
(2004). Plotted in grey are the points from the compilation of Hopkins
\& Beacom (2006, hereafter HB06). Our radio-derived SFRs do not come
from a flux-limited sample and thus can, in principle, span the full
radio luminosity function (LF) at all redshifts. However, the
$K$-selected sample we are stacking does lose lower $K$ luminosity
sources at higher redshifts. If the radio-derived SFR and $K$-band
luminosity are correlated then we will have an effective `luminosity
limit' to our radio stacks, as a function of redshift, which will mean
that we will be missing some fraction of the SFRD at higher
redshifts. We have checked this by stacking in bins of absolute $K$
mag ($M_K$) and the results are shown in Fig.~\ref{sfrKabsF}. We
stacked in two redshift bins ($z<1$ and $z>1$) to minimise any
redshift-dependent trends. The fits in both cases show ${\rm SFR}
\propto L_{K}^{1.01-1.05}$ which, within the uncertainties, is
consistent with a linear relationship.

We will now attempt to correct for the fraction of the radio
luminosity density lost from the faint end of the $K$-band LF. The
deepest wide-area study of the evolution of the $K$-band LF is
provided by the UDS (Cirasuolo et al.\ 2008) who present Schechter
function fits as a function of redshift. The integrated luminosity in
$K$ from infinity to some fixed lower luminosity ($L$) is given by:

\[
\rho_{L}(K) = \phi_{\ast} L_{\ast} \Gamma(\alpha+2, L/L_{\ast})
\]

\noindent where $\phi_{\ast}$ is the normalisation, $L_{\ast}$ is the
characteristic luminosity in $K$, $\alpha$ is the faint-end slope and
$\Gamma(\alpha+2, L/L_{\ast})$ is the incomplete Gamma function.

The correction we want to make ($C$) for each redshift bin is the
ratio of $\rho_{L}(K)$ integrated to the lowest observed luminosity at
that redshift -- the cut-off luminosity ($L_{\rm cut}$), to
$\rho_{L}(K)$ integrated to $L_{\rm min}$, the lowest observed
luminosity at $z=0$. We used the evolving $L_{\ast}$ from Cirasuolo et
al.\ (2008), with a fixed faint-end slope and calculated $L_{\rm cut}$
for each redshift bin. Thus

\[
C = \frac{\rho_{L}(K)_{\rm tot}}{\rho_{L}(K)_{\rm
obs}}=\frac{\Gamma(\alpha+2, L_{\rm min}/L_{\ast})}{\Gamma(\alpha+2,
L_{\rm cut}/L_{\ast})}
\]

\noindent The radio SFRD points were corrected by the factor $C$ and are shown
as solid circles on Fig.~\ref{SFRD}.  The uncorrected values are also
shown as open diamonds in Fig.~\ref{SFRD} for comparison.

The radio-derived SFRD is consistent with the multi-wavelength
compilation of HB06 out to $z\sim 2$. After this, the radio-derived
points appear to drop more steeply than the other data, even after
correction for the missing faint end part of the $K$-band LF.  This is
the highest redshift study of star formation in the radio to date and
it is not clear why the shape of the radio SFRD should be different to
that derived in the optical waveband. One possibility is that the
radio breaks down as a tracer of SFR at high redshift. Carilli \& Yun
(1999) suggest that radio emission could be quenched at high redshift
by inverse Compton scattering from the microwave background, but it is
believed that this should only be a serious effect at $z>6$. So far,
no evolution has been observed in the radio/FIR correlation out to
$z\sim 3$ (Kovacs et al.\ 2006; Ibar et al.\ 2008) but these studies
focus on samples with much higher $L_{\rm 1,400}$ and $L_{\rm
FIR}$. Our LF corrections may also be underestimated if the faint end
slope of the $K$-band LF undergoes evolution with redshift. The
optical/UV points at $z>2$ are also subject to substantial correction
for the effects of dust, which may not be appropriate at these
redshifts. Thus, we have no reason to suspect that the radio-derived
values are less valid than those derived at shorter wavelengths, and
some reason to suppose that they are rather more trustworthy.


We can also compare the radio-derived SFRD determined using the two
SFR relationships discussed earlier. Fig.~\ref{SFRD}a shows the
conversion of 1,400-MHz luminosity to SFR from Condon (1992) while
Fig.~\ref{SFRD}b shows that of Bell (2003). Both SFR calibration
methods straddle the points from HB06, with Condon's a little high,
and Bell's a little low. In Fig.~\ref{SFRD}, we also highlight as red
circles the literature points derived from other 1,400-MHz
surveys. The stacked points are in very good agreement with these
until $z\sim 1-2$. At the highest redshifts, the literature radio
studies are making very large corrections for the LF (which is not
measured at radio wavelengths at these distances) whereas our
corrections are more modest and are based on measured $K$-band
LFs. For comparison, our maximum LF correction is a factor 2$\times$
at $z\sim 4$, compared to a correction of 6$\times$ at $z\sim 2.5$ for
Seymour et al.\ (2008).

The lowest redshift point, at $z<0.2$, has been corrected (to
integrated and for {\sc clean}-to-dirty) by a factor 2.61, as
described in \S\ref{stack}. Without this differential correction for
the lowest redshift bin, this point would be very low compared to the
literature points from HB06. It is still noticeably lower than the
higher redshift points in Fig.~\ref{SFRD}a based on the Condon SFR
conversion, though using the Bell SFR gives better agreement. Given
our reliance on photometric redshifts, where the magnitude of $\delta
z/z$ is of order unity for $z<0.2$, it is possible that uncertainties
in redshift are producing this decrement.  Until more spectroscopic
redshifts are available to help calibrate the photometric redshifts,
we cannot address this issue adequately and so will not discuss
sources at $z<0.2$ further.

The general agreement between the two SFR estimators and the
literature work further supports the mounting evidence that AGN are
not strongly contributing to the globally-averaged radio emission of
these stacked samples.  Overall, at this point, we feel there is
nothing to choose between the two SFR conversions and so will continue
to use both in the following section.

\subsection{Specific star-formation rates}

We can also look at the average SFR per unit stellar mass, defined as
the SSFR, as a function of stellar mass and redshift. SSFR is often
used as a measure of the star-formation efficiency of a galaxy, since
it provides information about the fraction of a galaxy's mass which
could be converted into stars in a given time. A higher SSFR means a
galaxy will increase its mass by a greater fraction in a given time
than a lower SSFR.

\begin{figure*}
\includegraphics[width=8.7cm]{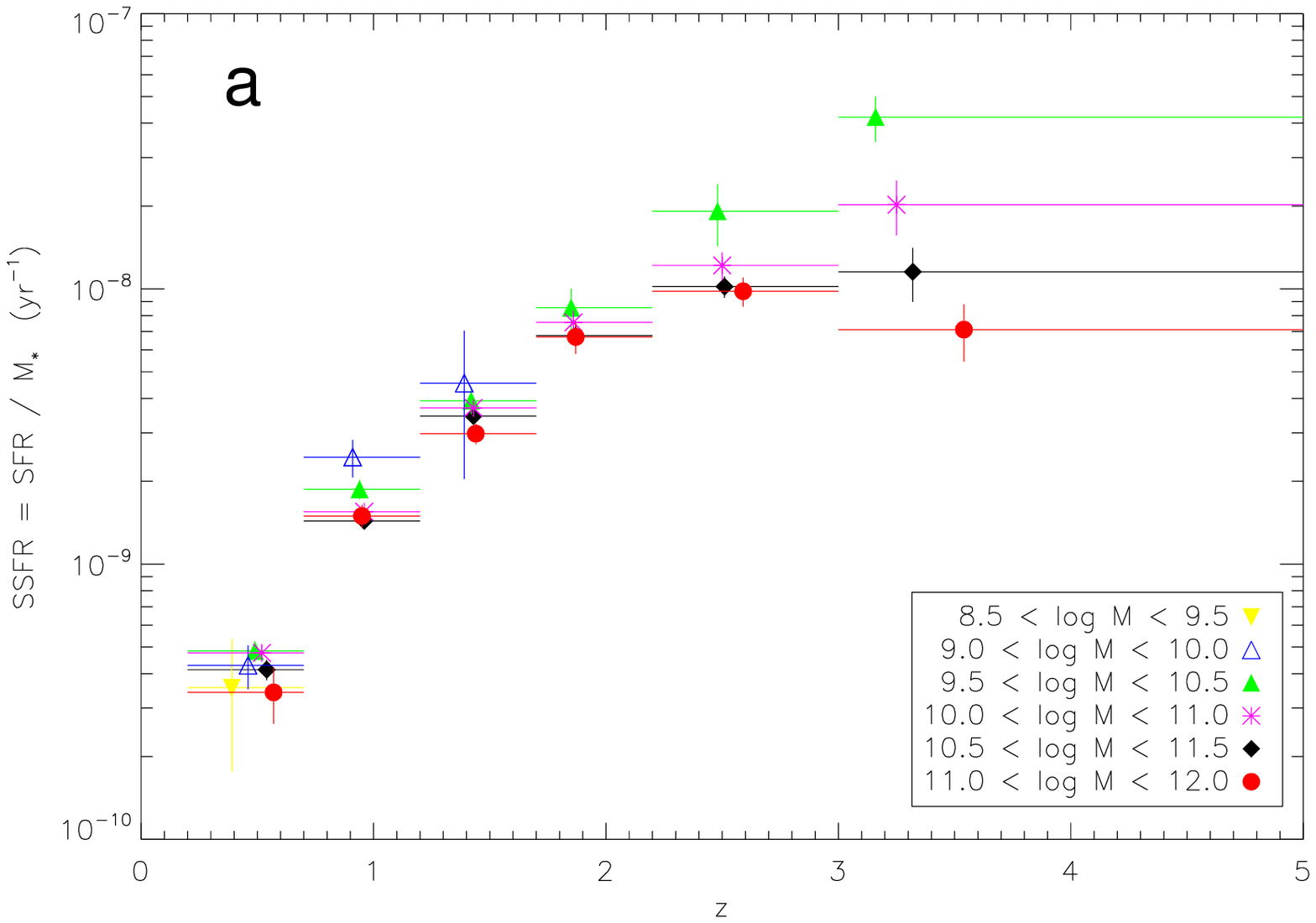}
\includegraphics[width=8.7cm]{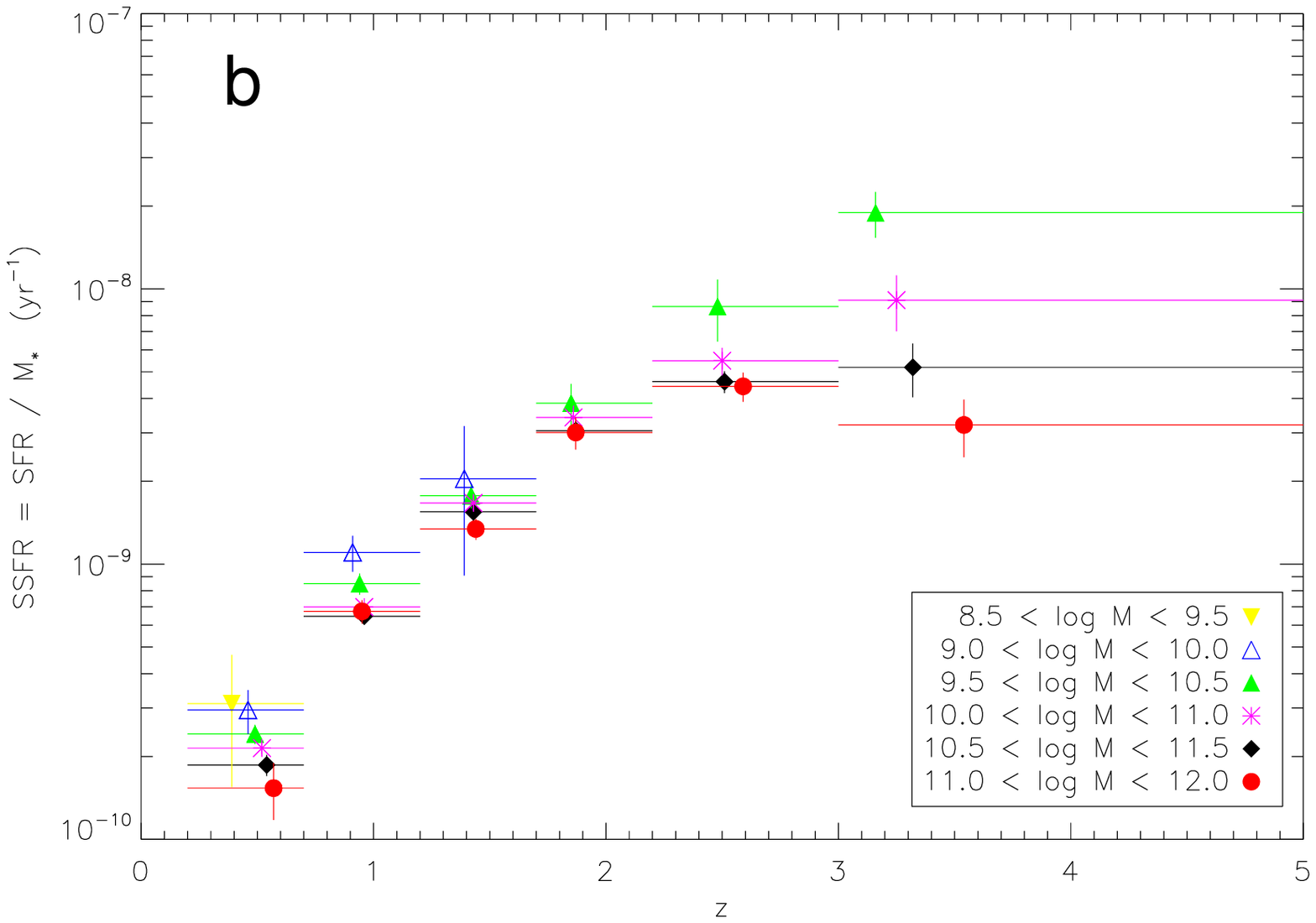}
\caption{\label{SSFR_z_all} SSFR (SFR/$M_{\rm stellar}$) as a
function of redshift for the UDS galaxies. Error bars are based on the
1$\sigma$ confidence interval on SSFR, as measured directly from the
distribution of values in the stack, with no account taken of possible
systematics from determination of $M_{\rm stellar}$ via $K$ mag. (a)
shows SSFR using the Condon SFR conversion and (b) is the
Bell SFR conversion. Various symbols represent different $M_{\rm stellar}$
bins as explained in the legend.}
\end{figure*}

The results for the full $K$-selected sample are shown in
Fig.~\ref{SSFR_z_all}, where Fig.~\ref{SSFR_z_all}a uses Condon's SFR
and Fig.~\ref{SSFR_z_all}b uses Bell's SFR. We see a strong trend of
increasing SSFR with redshift for all bins of stellar mass. There is
no difference in the evolution of SSFR with redshift for different
mass bins out to $z\sim 2$, after which the high-mass bins tend to
flatten off compared to the low-mass bins.  There is also a trend that
the least massive galaxies at any redshift have the highest SSFR
(becoming more pronounced with increasing redshift). The difference
between Bell's and Condon's SFRs are clear in two ways. First, the
SSFRs are higher overall for the Condon SFR estimator and, second, the
trends in the lowest redshift bin are significantly different. This is
because the low-redshift bin contains a number of mass bins where the
radio luminosity is below $L_c$ in the Bell (2003) relationship. As a
result, the Bell (2003) conversion boosts these low $L_{\rm 1,400}$
points to account for his finding that low-luminosity galaxies are
deficient in non-thermal radio flux for a given
SFR. Fig.~\ref{SSFR_z_all}b therefore shows a trend for the lowest
mass galaxies to have the highest SSFR in the lowest redshift bin,
whereas Fig.~\ref{SSFR_z_all}a (Condon SFR) does not.

A large volume of literature on SSFR has been produced since recent
advances in deriving robust stellar masses for large galaxy samples
through deep and wide $K$-band and {\em Spitzer} imaging (Feulner et
al.\ 2005, 2007; Caputi et al.\ 2006; Zheng et al.\ 2007; Daddi et
al.\ 2007; Elbaz et al.\ 2007; Iglesias-Paramo et al.\ 2007;
Perez-Gonzalez et al.\ 2007; Salim et al.\ 2007; Martin et al.\ 2007;
Buat et al.\ 2007). These investigations all report two similar
trends, though the exact values tend to differ depending on the
conversions to $M_{\rm stellar}$ and the method used to determine
SFR. Firstly, they find that SSFR increases with redshift for all
masses of galaxy -- high-redshift galaxies were forming stars not only
at a higher absolute rate than low-redshift galaxies, but also at a
higher rate compared to their stellar mass. Secondly, SSFR studies in
the literature also show that the highest SSFR at any given redshift
is displayed by the lowest mass galaxies, with many studies finding
that SSFR has dropped more rapidly since $z\sim 2$ in the most massive
galaxies (Feulner et al.\ 2005; Perez-Gonzalez et al.\ 2007). Many
authors have claimed that this difference in the SSFR behaviour at
high and low masses is akin to differential evolution -- the so-called
`downsizing', discovered originally via a $K<22$ ($B<24.4, I<23$)
survey and described as a `smooth decrease in the characteristic
luminosity of galaxies dominated by star formation' (Cowie et al.\
1996).

\begin{figure*}

a) SSFR using the Condon SFR conversion.

\includegraphics[width=14.5cm]{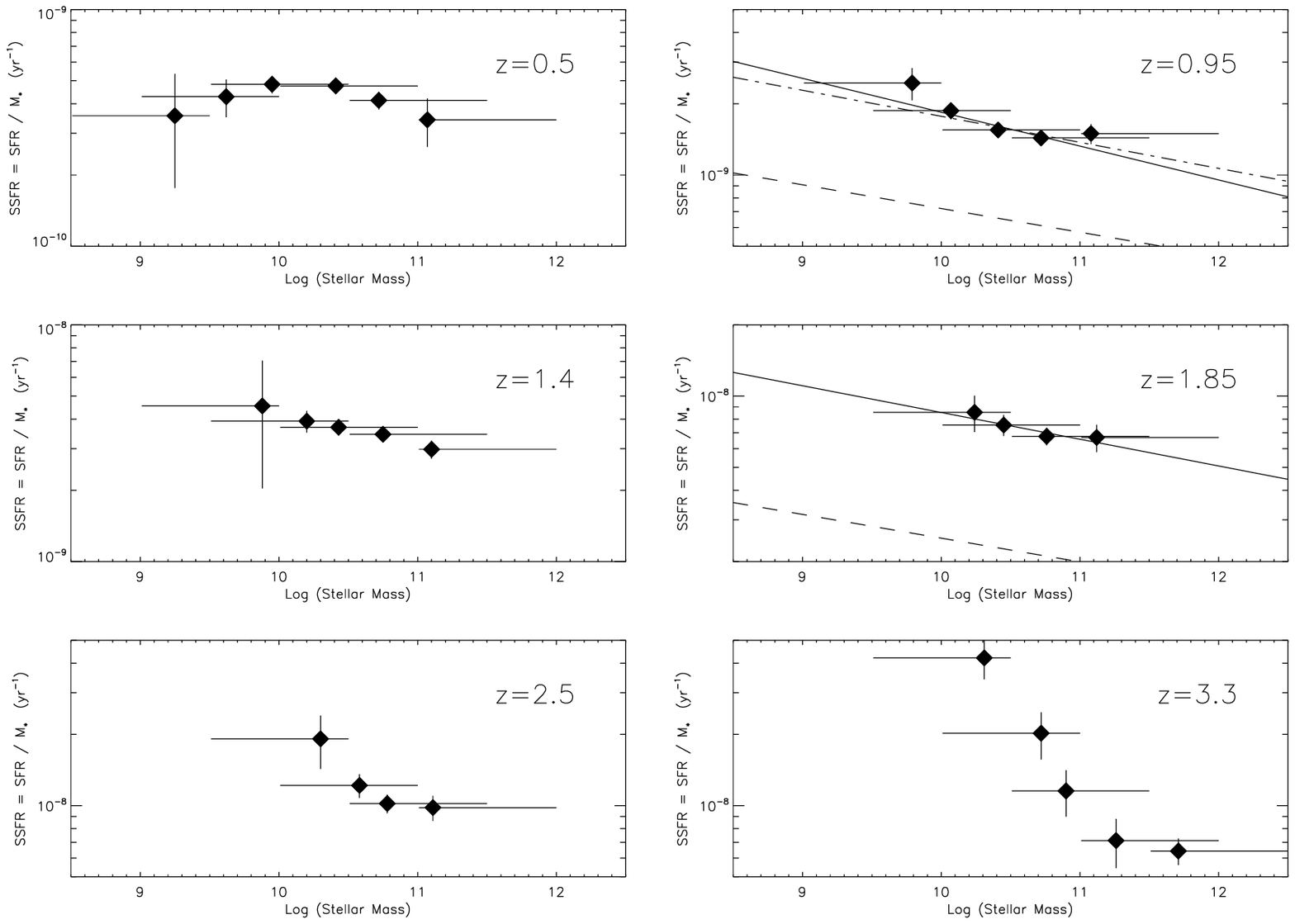}

b) SSFR using the Bell SFR conversion.

\includegraphics[width=14.5cm]{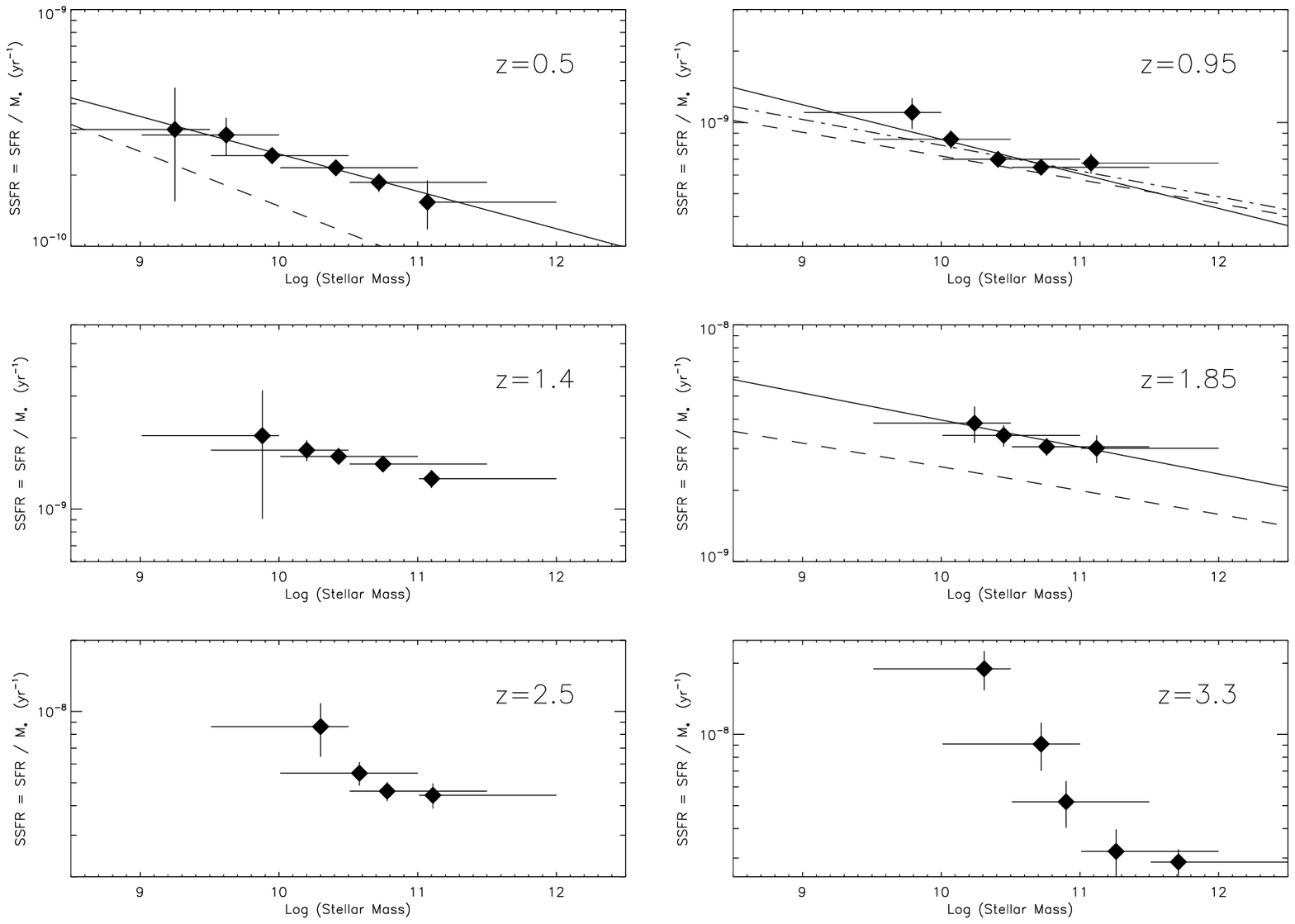}

\caption{SSFR as a function of stellar mass for various redshift
bins. The dynamic range of the sub-panels is the same, so slopes are
directly comparable. Masses plotted are the median mass from the bins
in Fig.~\ref{SSFR_z_all}. The panels for $z\sim 0.5$, $z\sim 1$ and
$z\sim 2$ are compared to the results of Brinchmann et al. \ (2004),
Elbaz et al.\ (2007) and Daddi et al.\ (2007). The solid line shows
the fit to our data while dashed lines show fits from the
literature. The dash-dot line in the $z\sim 1$ panel is the fit
without the lowest-mass bin. This is in keeping with the mass range
fitted by Elbaz et al.\ (2007). The slopes are consistent at $z=1-2$,
but the normalisation is different due to the different SFR indicators
and $M_{\rm stellar}$ estimates. At $z=0.5$, the slope is intermediate
between that of Brinchmann et al.\ (2004) at $z=0$ and Elbaz et al.\
(2007) at $z=1$.}
\label{SSFR_mass}
\end{figure*}

We will now make a detailed comparison between our results and
those in the literature and discuss possible reasons for the
differences.

Fig.~\ref{SSFR_mass} shows in more detail the relationship between
SSFR and $M_{\rm stellar}$. Fig.~\ref{SSFR_mass}a uses the Condon SFR
and Fig.~\ref{SSFR_mass}b uses the Bell SFR. From this we can notice
two things. First, most panels show a weak relationship between SSFR
and $M_{\rm stellar}$, where the steepness of the correlation appears
to increase with redshift. Second, we note the difference in
appearance of the lowest redshift bin when using the different SFR
conversions. With the Condon SFR, there is no apparent trend while for
the Bell SFR, there is a significant correlation. Given the number of
studies finding such a relationship at low redshift, and that we also
find a similar trend at higher redshifts, we feel that this is
tentative evidence supporting the Bell SFR estimation.
 
We have over-plotted comparisons with other literature where
available. The studies by Daddi et al.\ (2007) of BzK galaxies at
$z=2$, and by Elbaz et al.\ (2007) of sources at $z=1$ selected using
24-$\mu$m data in the Great Observatories Origins Deep Survey (GOODS),
are over-plotted on the $z=0.95$ and $z=1.85$ panels of
Fig.~\ref{SSFR_mass}. They find a correlation between SFR and $M_{\rm
stellar}$ which is almost linear (SFR $\propto M_{\rm stellar}^{0.9}$)
leading to a shallow dependence of SSFR on mass. The study of $z=0$
galaxies from the SDSS by Brinchmann et al.\ (2004) finds a shallower
slope of ${\rm SFR} \propto M_{\rm stellar}^{0.77}$. The Brinchmann et
al.\ trend is overplotted at $z=0.5$ to show that the slope of the
relationship we find is intermediate between that at $z=0$ and $z=1$.

The rather weak trends of SSFR versus $M_{\rm stellar}$ found here are
at some odds with other findings in the literature. Some studies find
that the differences in SSFR at $z=0.6-1$ over the mass range $8.5 <
\log\, M_{\rm stellar} <11.5$ to be a factor $10-32$ compared to our
differences of a factor $1.6-2$ (Feulner et al.\ 2005; Caputi et al.\
2006; Zheng et al.\ 2007; P\'{e}rez-Gonz\'{a}lez et al.\ 2007). In our
highest redshift bins ($z>2$) we can see a steepening of the
relationship between SSFR and $M_{\rm stellar}$, however these are not
well-fitted by a power law. Extrapolating the linear portion of the
$z=3.3$ panel, we find that the change in SSFR over the range $8.5 <
\log\, M_{\rm stellar} <11.5$ at $z\sim 3.3$ is a factor $\sim 25$ --
closer to the values seen at lower redshift in the other studies.

Most of the surveys reporting these steeper trends have two things in
common. First, they are selected in the optical and use $K$ or {\em
Spitzer} data to estimate $M_{\rm stellar}$. Second, they rely on
detections of galaxies in one or more bands, from which the SFR is
then directly inferred (e.g.\ UV, 24\,$\mu$m). The rest-frame UV is
the most sensitive to low levels of SFR and thus the method by which
most of the low-mass bins have been determined. For a low-mass galaxy
with a low SSFR to be part of such a sample, the sensitivity to SFR in
the survey must be very high and there is therefore a potent selection
bias: low-mass galaxies with higher-than-average SSFRs are included
preferentially in such samples. This was noted by Feulner et al.\
(2007), who investigated the effects of survey wavelength on the slope
of the SSFR vs $M_{\rm stellar}$ relation at various redshifts. They
found that $B$-band surveys select via SFR, effectively, while
$K$-band surveys select via stellar mass (for $z<1.5$, at
least). $B$-band selection produces greater correlation between SSFR
and $M_{\rm stellar}$. This selection band bias most strongly affects
low-mass galaxies. Elbaz et al.\ (2007) and Zheng et al.\ (2007) also
noted that optical selection for spectroscopic completeness leads to a
loss of red (low-SFR) low-mass galaxies, pushing up the apparent SSFR
in the lowest-mass bins, which suggests at least one pragmatic
argument in favour of using high-quality photometric redshifts.

Our sample is based on selection via an extremely deep $K$-band survey
(where selection is not strongly dependent on current SFR at $z<1.5$)
and the current SFR is determined by radio stacking (rather than via a
flux-limited radio sample). The weak dependence of SSFR on $M_{\rm
stellar}$ at $z<2$ found here -- and the similar evolution of SSFR
with redshift across all masses -- could be explained by our less
biased selection. At $z>2$, $K$ is shifting into the rest-frame
optical, making our selection more SFR-based than $M_{\rm
stellar}$-based. As predicted, this causes a stronger apparent trend
of SSFR with $M_{\rm stellar}$ because high-SSFR objects are
preferentially selected near the limit of the survey.

In summary, many independent studies find similar trends, with SSFR
increasing with redshift and decreasing with stellar mass. However,
the strength of these relationships varies considerably and is likely
to be due to a complicated combination of sample-selection criteria,
particularly wavelength and depth. Many authors make strong statements
about downsizing, based on the evolution of SSFR with redshift in
different stellar mass bins, and on the oft-strong correlation of SSFR
with stellar mass. {\em We would urge caution before over-interpreting
this type of plot as it is influenced strongly by selection biases.}
Given that our exploration of SFR versus absolute $K$ mag produced a
linear correlation, with $\rm SFR \propto L_K^{1.01-1.05}$, it is the
conversion of $K$ mag to stellar mass which introduces the slight
non-linearity in SFR versus $M_{\rm stellar}$ ($\rm SFR \propto M_{\rm
stellar}^{0.9}$). This, in turn, produces the shallow observed
dependence of SSFR on $M_{\rm stellar}$. Thus, the conversion to
stellar mass can also introduce systematic effects, producing trends
in SSFR with $M_{\rm stellar}$.

The upcoming addition of extremely deep $U$, $H$ and 3.6--24-$\mu$m
{\em Spitzer} imaging, as well as several thousand more spectroscopic
redshifts, the stellar masses and redshifts of the UDS sample will
improve markedly. At that time, it will be interesting to test whether
these trends remain.

\subsection{Specific star-formation rates in BzK galaxies}

\begin{figure}
\includegraphics[width=8.6cm]{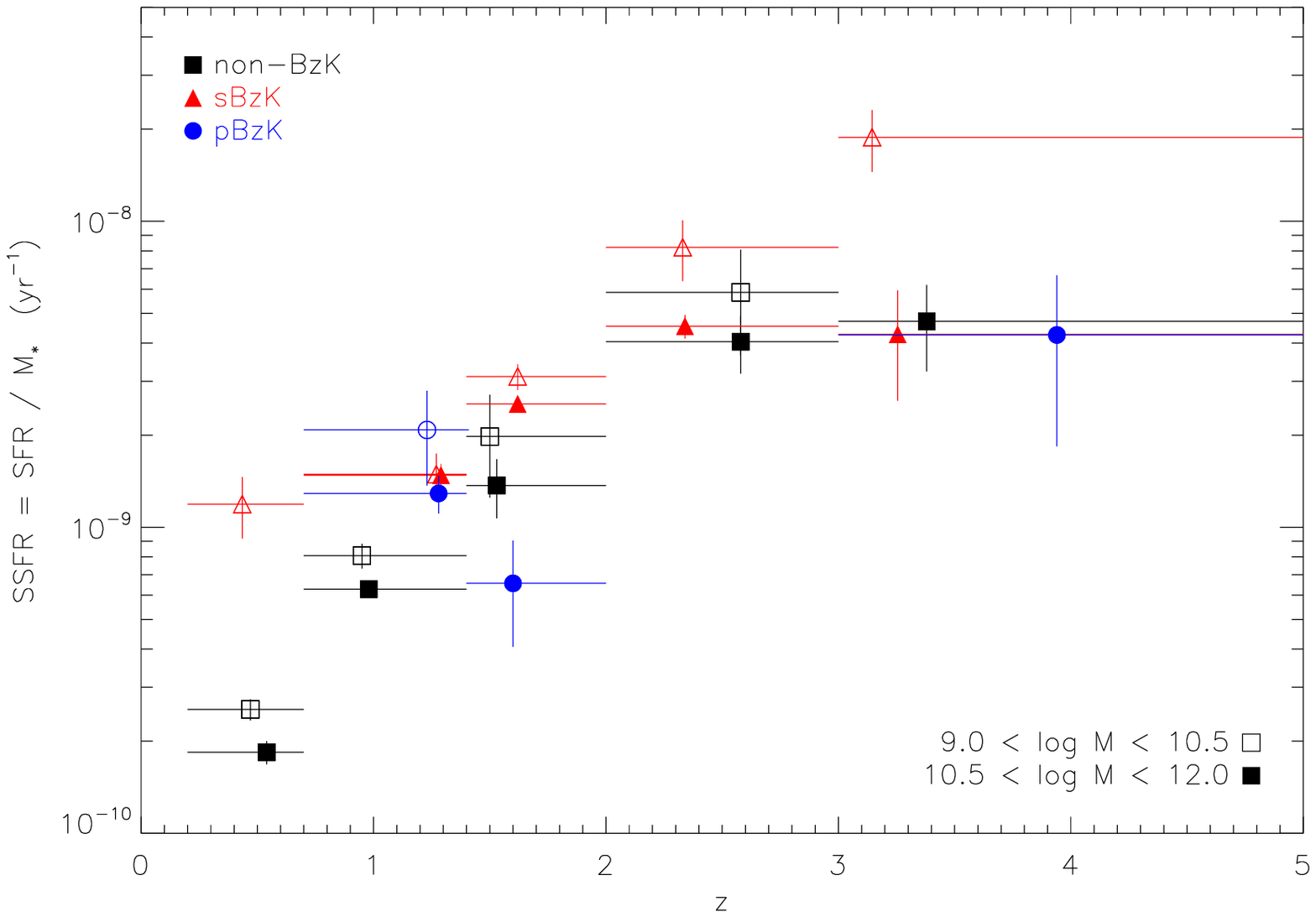}
\caption{\label{SSFR_bzk} SSFR using the Bell's SFR conversion with
redshift and mass for sBzk (red triangles), pBzK (blue circles) and
non-BzK galaxies (black squares). Objects with log\,$M_{\rm stellar}=
9.0-10.5$ ($10.5-12.0$) are denoted by open (filled) symbols.}
\end{figure}

We have further split the samples of sBzK, pBzK and non-BzK galaxies
into bins of redshift and stellar mass to investigate their SSFR
trends. This is shown in Fig.~\ref{SSFR_bzk} with different colours
and symbols denoting the different samples and open and filled symbols
used to denote the mass ranges. Here we see a similar trend to
Fig.~\ref{SSFR_z_all}, with SSFR increasing with redshift. The sBzK
galaxies tend to have slightly higher SSFRs for a given mass than the
non-BzK galaxies, consistent with the idea that they are a strongly
star-forming population. In the lowest-redshift bin ($0.2<z<0.7$),
sBzK galaxies are forming stars more actively than non-BzK galaxies in
the same redshift and mass range. The subset of sBzK galaxies at low
redshift have very blue colours and are fainter in $K$ than non-BzK
galaxies in the same redshift interval. Kolmogorov-Smirnov (K-S) tests
show that the radio flux densities, SSFRs and $K$ mag distributions
are significantly different from the main low-redshift non-BzK galaxy
sample at probabilities of $0.015,\ 1.3\times 10^{-9}$ and $9.2\times
10^{-13}$ respectively. The flat, blue spectra of these galaxies makes it very
difficult to assign accurate photometric redshifts and, as they
represent only a small fraction of the galaxies at this redshift
($\sim$1--2 per cent) they could be photometric-redshift outliers. If
their redshifts were increased to $z\sim 1-1.4$, their SSFRs would not
be remarkable compared to other galaxies at this redshift. If they
truly do lie at low redshift, they represent a population of low-mass,
highly star-forming galaxies, quite unlike the general low-redshift
population. Confirmation of their nature must await spectroscopy.

The pBzK galaxies behave quite differently. Their SSFR drops from
$0.7<z<1.4$ to $1.4<z<2.0$ and then appears to be consistent again
with the other samples at very high redshifts. We have already noted
changes in pBzK properties at $z=1.3-1.4$ from their radio spectral
indices (\S\ref{GMRT}), the decrease in their radio flux density and
the suggestion of a similar decrease in their submm flux density. This
argues that the BzK diagram alone is not discriminating effectively
between passive and star-forming galaxies -- accurate redshifts are
also required. The persistent (but lower level) radio flux density at
$z>1.4$ could be due either to low-level SFG contamination or to
radio-quiet AGN activity. The radio spectral indices and submm
stacking also suggest a change in the dominant pBzK population at
$z\sim 1.4$, but our submm and 610-MHz observations are not deep
enough to allow us to conclude which of the above scenarios is
responsible for the low-level 1,400-MHz emission at $z>1.4$.

Future analysis of the larger SHADES 1.1-mm AzTEC image, newly
acquired {\em Spitzer\/} data, alongside the zUDS spectroscopic
programme, should determine the true nature of pBzK galaxies.

\section{Conclusions}

We have stacked several $K$-selected populations into a deep 1,400-MHz
mosaic in order to investigate their star-formation history.  There is
much dispute over the relative contribution from star formation and
AGN to the radio populations at sub-mJy levels and our knowledge is
almost non-existent at flux levels below those where individual
galaxies are usually detected ($<$100\,$\mu$Jy). We certainly cannot
distinguish unambiguously between star-formation- and AGN-related
emission. However, their spectral indices, submm flux densities and
the good agreement between SFRDs derived in the radio and that
determined at other wavelengths argues that radio-quiet AGN do not
make an important contribution to the average radio properties of our
$K$-selected galaxies at these flux densities (5--20$\, \mu$Jy).

We find a strong relationship between stacked radio flux density and
apparent $K$ mag. We also find SFR to be a strong and almost linear
function of absolute $K$ mag and stellar mass.

Both sBzK and pBzK galaxies are detected robustly in the radio
stack. The stacked radio flux density of the sBzK galaxies is roughly
constant with redshift, inferring a strong evolution in SFR.

Our photometric redshifts suggest that a significant fraction
($\sim$30--40 per cent) of galaxies occupying the sBzK and pBzK
regions of the BzK diagram lie at $z<1.4$. The similarity between sBzK
and $z<1.4$ pBzK galaxies -- in terms of their radio- and submm-derived SFR
and SSFR -- leads us to suggest that the BzK diagram
alone is not sufficient to select passive high-redshift galaxies. Only
pBzK galaxies at $z>1.4$ seem consistent with this designation, those
at $z<1.4$ are likely to be star forming galaxies similar in nature to sBzK.

The pBzK galaxies suffer a dramatic reduction in radio flux density at
$z>1.4$. Their weak radio flux density at $z>1.4$ suggests either a
persistent, low-level of contamination by SFGs, or that the radio
emission for these objects is powered by radio-quiet AGN. Deeper
610-MHz and submm data are required to determine which is the case.

The variation of radio-derived SFR density with redshift agrees well
with that determined at other wavelengths, for $z=0.2-2.0$. This
suggests that the contribution from AGN-related radio emission is
small. At $z\ge 2$, the radio-derived SFR density declines due to our
inability to fully sample the $K$-band luminosity function at these
redshifts. After correcting for this, using the evolving $K$-band UDS
LFs, we still see a downturn in the radio-derived SFRD at high
redshifts which is not apparent in the other multi-wavelength
data. The reason for this is not clear but we suggest three
possibilities: (a) a changing slope with redshift for the $K$-band LF,
which would mean our LF corrections have been underestimated; (b) dust
corrections to the optical/UV data are too high; (c) the radio
emission is not tracing star formation efficiently at high redshift,
possibly due to evolution of the magnetic field properties or a change
in the mode of star formation.

We find that the SSFR (SFR/$M_{\rm stellar}$) is only weakly dependent
on stellar mass, with SSFR decreasing as $M_{\rm stellar}$
increases. The observed correlation of SSFR with stellar mass is
consistent with the shallower determinations in the literature, such
as Daddi et al.\ (2007) at $z\sim 2$ and Elbaz et al.\ (2007) at
$z\sim 1$. It is much less steep than many optical-/UV-based studies
and we argue that these differences are largely due to selection
biases present in UV- and optically-selected samples. We also find a
very strong trend of increasing SSFR with redshift, stronger than
reported elsewhere, with a flattening at $z\gs 2$. This is also where
the difference between high- and low-mass bins becomes greatest and
where $K$ begins to sample the rest-frame optical waveband, meaning
that sample selection becomes more influenced by star formation and
less by stellar mass.

Comparing the radio-derived SFR conversions from Bell (2003) and
Condon (1992), the former provides a better match to the observed
trends in SSFR versus stellar mass in the lowest mass bins, and also
in reproducing the low-redshift SFRD seen in other wavebands.

In summary, the UDS has produced a statistically powerful sample
of $K$-selected galaxies, including several thousand selected by
colour to lie at high redshift. Stacking these samples into a deep
1,400-MHz radio image has enabled us to determine their radio
properties at flux densities an order of magnitude fainter than has
previously been possible with flux-limited radio surveys. The
pseudo-stellar-mass selection achieved in the $K$ band -- combined
with a probe of star formation which is not limited to the rarest,
brightest sources -- has allowed us to follow the SFRD out to $z\sim
4$ and to investigate the evolution of SSFR in galaxies with redshift
in a relatively unbiased way.

\section*{Acknowledgements}

We thank the anonymous referee for helpful comments on the paper.  We
would like to thank Stephen Serjeant for useful discussions. OA would
like to ackowledge the support of the Royal Society. SF, AJM, RM and
MC acknowledge support from STFC.

\end{document}